\documentclass[11pt,twoside]{article}
\usepackage{fleqn,espcrc1}
\input{epsf}

\title{\large \bf Modulated Amplitude Waves and Defect Formation
in the\\ One-Dimensional Complex Ginzburg-Landau Equation}

\author{Lutz Brusch\address[1]{ Max-Planck-Institut f\"ur Physik
  komplexer Systeme, N\"othnitzer Stra{\ss}e 38, D-01187 Dresden,
  Germany},
Alessandro Torcini\address[2]{ Dipartimento di Fisica, Universita'
  ``La Sapienza'', P.le A. Moro 2, I-00185 Roma, Italy and\\ Istituto
  Nazionale di Fisica della Materia, Unit\`a di Firenze, Largo Enrico
  Fermi 2, I-50125 Firenze, Italy}, 
Martin van Hecke\addressmark[1]\address[3]{ Center for Chaos and
  Turbulence Studies, The Niels Bohr Institute, Blegdamsvej 17, 2100
  Copenhagen, Denmark and\\ Kamerlingh Onnes Laboratory, Leiden
  University, Niels Bohrweg 2, 2333CA Leiden, The Netherlands},
Mart{\'\i}n G. Zimmermann\address[4]{ Instituto Mediterr\'aneo de
  Estudios Avanzados, IMEDEA (CSIC-UIB), E-07071 Palma de Mallorca,
  Spain and\\Departamento de F{\'\i}sica, FCEN-Universidad de Buenos
  Aires, Pab. I Ciudad Universitaria, 1428 Buenos Aires, Argentina}
and
Markus B\"ar\addressmark[1]}
 
\begin{document}
 
\maketitle 

\begin{abstract}

\noindent
{\bf Abstract}

\noindent
The transition from phase chaos to defect chaos in the complex
Ginzburg-Landau equation (CGLE) is related to saddle-node bifurcations
of modulated amplitude waves (MAWs). First,
the spatial period $P$ of MAWs is shown to be limited by a maximum
$P_{SN}$ which depends on the CGLE coefficients; MAW-like structures
with period larger than $P_{SN}$ evolve to defects.  Second, slowly
evolving near-MAWs with average phase gradients $\nu \approx 0$ and
various periods occur naturally in phase chaotic states of the CGLE.
As a measure for these periods, we study the distributions of spacings
$p$ between neighboring peaks of the phase gradient.  A systematic
comparison of $p$ and $P_{SN}$ as a function of coefficients of the
CGLE shows that defects are generated at locations where $p$ becomes
larger than $P_{SN}$.  In other words, MAWs with period $P_{SN}$
represent ``critical nuclei'' for the formation of defects in phase
chaos and may trigger the transition to defect chaos.  Since rare
events where $p$ becomes sufficiently large to lead to defect
formation may only occur after a long transient, the coefficients
where the transition to defect chaos seems to occur depend on system
size and integration time.  We conjecture that in the regime where the
maximum period $P_{SN}$ has diverged, phase chaos persists in the
thermodynamic limit.



\end{abstract}

\section{Introduction}

The transition from {\it phase} to {\it defect chaos} for the one
dimensional complex Ginzburg-Landau equation (CGLE) was recently
related to the bifurcation properties of a family of coherent
structures called {\em modulated amplitude waves} (MAWs) \cite{maw}.
In this paper the relationship between MAWs and large scale chaos is
studied in detail, providing a comprehensive description of various
aspects of the CGLE chaotic dynamics.

When a spatially extended system is driven sufficiently far away from
equilibrium, patterns can eventually form~\cite{kura,CH}.
In many cases these patterns show an erratic behavior in space and
time: such behavior is commonly referred to as 
{\em spatiotemporal chaos}~\cite{kura,CH,KS,review}. 
Examples of extended systems displaying such chaotic dynamics 
in one spatial dimension include: heated
wire convection~\cite{wire}, printers instability and film drag 
experiments~\cite{film}, eutectic growth~\cite{eutectic}, 
binary convection~\cite{binary}, sidewall convection~\cite{sidewall}, 
the far field of spiral waves in the Belousov-Zhabotinsky
reaction~\cite{chem}, the Taylor-Dean system~\cite{dean},
hydrothermal~\cite{SupEck} and internal~\cite{velarde} waves excited by the 
Marangoni effect and the oscillatory instability of a Rayleigh-B\'enard
convection pattern~\cite{janiaud}.

Near the pattern forming threshold, the dynamics of such systems can often be
described by so-called amplitude equations. 
When the pattern forming bifurcation 
from the homogeneous state is a forward Hopf bifurcation, the
appropriate amplitude equation is the CGLE
\cite{kura,CH}, which in one spatial dimension reads as:
\begin{equation} 
  \partial_{t} A = A + (1+ ic_1) \partial_{x}^{2} A - (1-i c_3) |A|^2 A
  ~, \label{cgle}
\end{equation}
where $c_1$ and $c_3$ are real coefficients and the field $A=A(x,t)$
has complex values.
 
For different choices of the coefficients
numerical investigations of the CGLE have revealed the existence of
various steady and spatiotemporally chaotic states
\cite{maw,kura,CH,KS,review,janiaud,pumir,chao1,bazhenov,chate,saka,egolf,miguel,torcini,saar,mvh,hegger}.
Many of these states appear to consist of individual structures with well
defined propagation and interaction properties.  It is thus tempting
to use these structures as building blocks for a better understanding 
of spatiotemporal chaos.
In this paper we will essentially follow such an approach. 

As a function of the coefficients $c_1$
and $c_3$, the CGLE (\ref{cgle}) can exhibit two qualitatively different
spatiotemporal chaotic states known as phase chaos 
(when the modulus $|A|$ is at any time bounded away from zero) 
and defect chaos (when $|A|$ can vanish leading to phase singularities).
It is under dispute whether the 
transition from phase to defect chaos is sharp or not, and 
if a pure phase-chaotic, ({\it  i.e.,}
defect-free) state can persist in the thermodynamic limit
\cite{egolf,chate2d}. 
We will address these issues by suggesting a {\em mechanism} for the formation 
of defects related to the range of existence of MAWs.

\begin{figure} \vspace{0cm}
\epsfxsize=1.2\hsize \mbox{\hspace*{-.1 \hsize} \epsffile{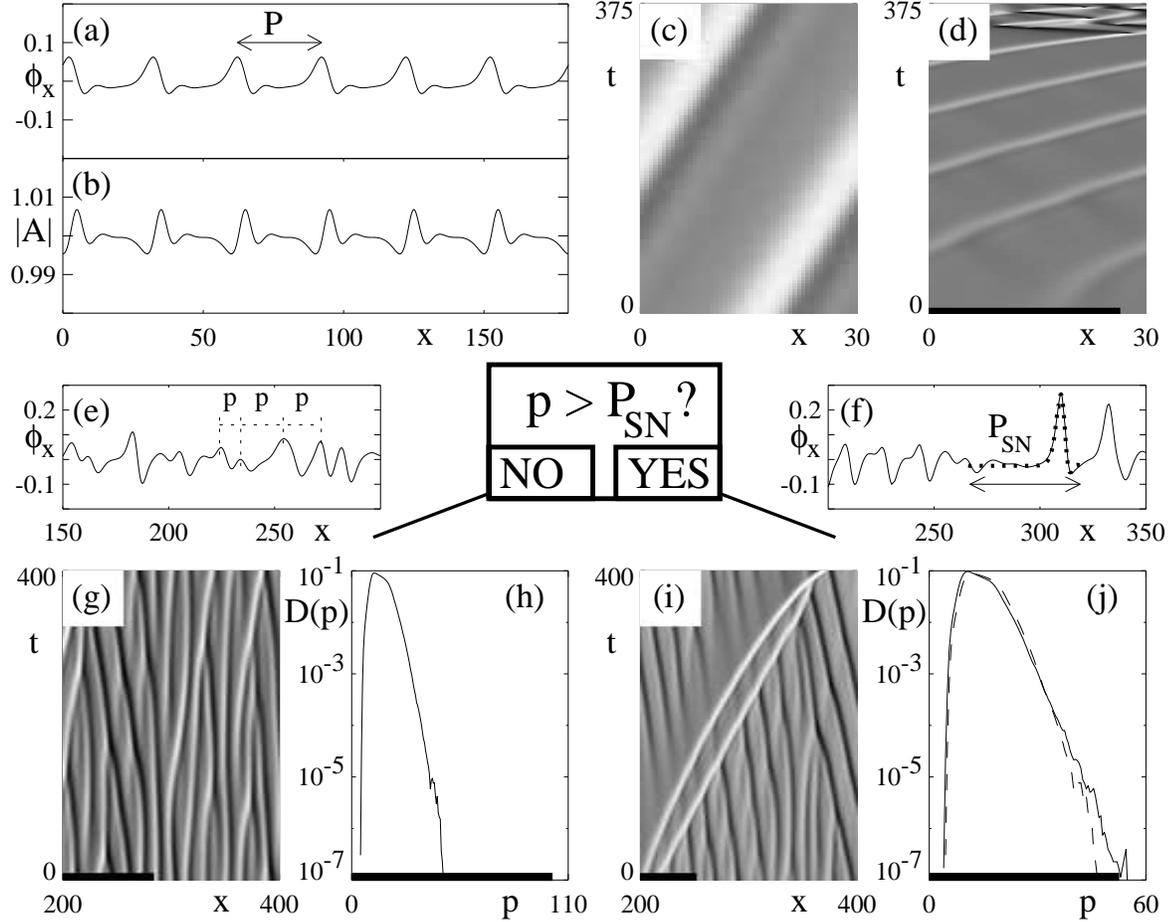} }
\vspace{-2.cm}
\caption[]{Summary of our main results which constitute a picture for
the formation of defects from phase chaotic states.  (a,b) Example of
a coherent structure: phase gradient and modulus of a period
$P\!=\!30$ MAW at $c_1\!=\!0.6, c_3\!=\!2$.  (c) Space time plot
showing the stable propagation of the MAW from (a,b) in a small system
of size $P$ with periodic boundary conditions. Subsequent space time
plots also show the phase gradient encoded in gray-scale (minima appear
dark, maxima bright). (d) The same MAW as initial condition creates
defects at $c_1=0.7, c_3=2$ where $P>P_{SN}=26.8$.  Black bars above
the x-axis denote the size of $P_{SN}$ specific to the parameters of
the panel.  (e,g,h) Large scale chaos at $c_1=0.63, c_3=2, L=512$. (e)
Snapshot of the phase gradient profile with individual inter-peak
spacings $p$. (g) Space time evolution of phase chaos and (h)
distribution $D(p)$ showing $p \ll P_{SN}$ and no defects. A transient
of $t\!\approx\!10^4$ is not shown.  (f,i,j) Large scale chaos at
$c_1=0.65, c_3=2, L=512$. (f) Snapshot of the phase gradient profile
$t=120$ before the first defect forms and the MAW (dotted, $P=P_{SN}$)
overlayed onto the long structure. (i) Transient phase chaos with a
fast and long structure traveling through the system which eventually
nucleates defect chaos at $t=400, x=360$ (a transient of
$t\!\approx\!10^4$ is not shown). A snapshot of this structure
was shown in (f). (j) The tail of the distribution of $p$
reaches $p>P_{SN}$ due to the long structure; this leads to the break
down of phase chaos.  The distribution $D(p)$ shown in (h) is also
reported (dashed line).  From the comparison of the two it is evident
that the distributions do not modify dramatically when $c_1$ is
increased, while $P_{SN}$ decreases noticeably.  }\label{fig0}
\end{figure}
\noindent

\begin{figure}[tbhp]
 \epsfxsize=.6\hsize \mbox{\hspace*{.15 \hsize} \epsffile{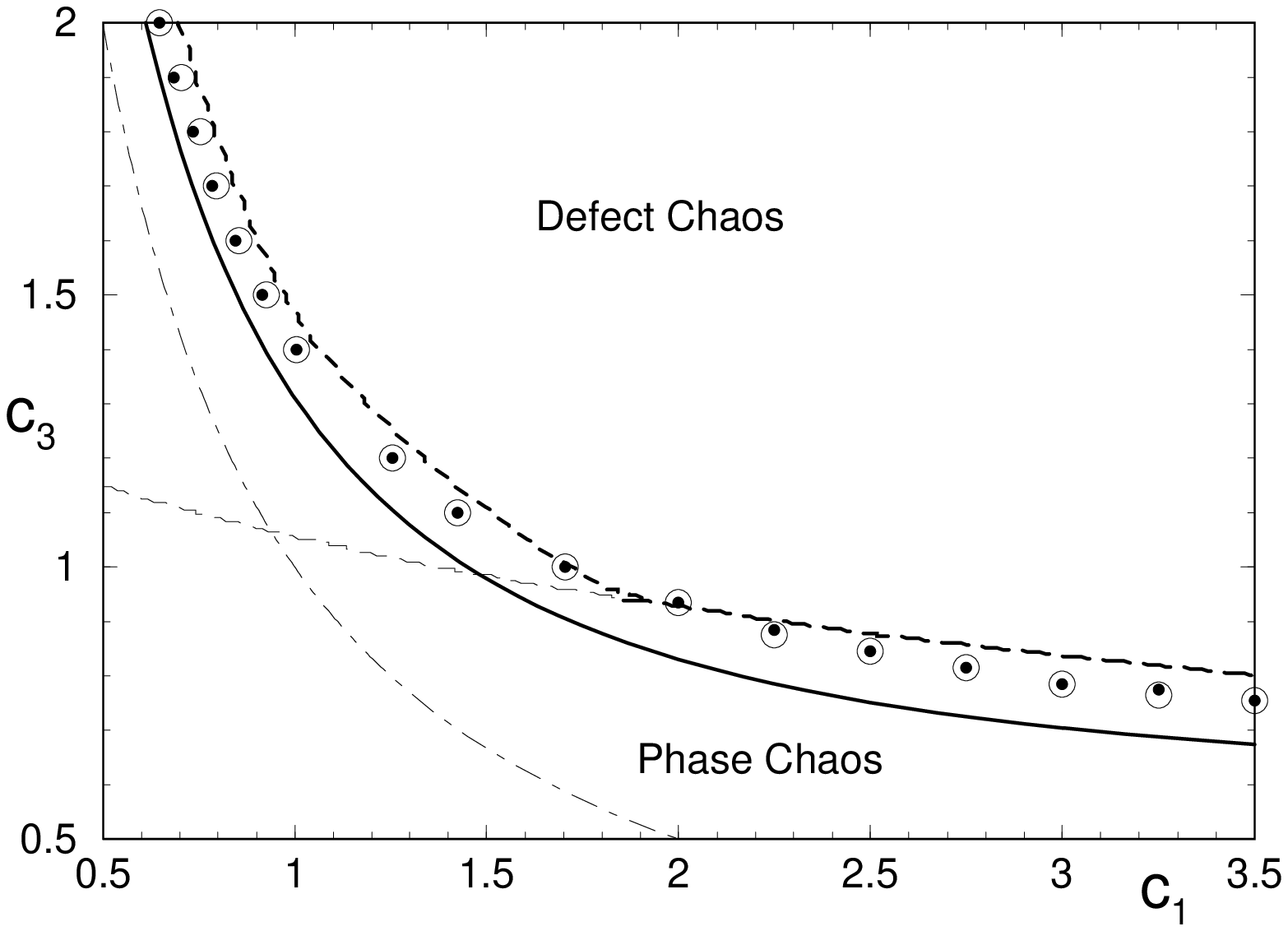} }
\vspace{-1cm}
\caption[]{ Phase diagram of the CGLE, showing the
Benjamin-Feir-Newell curve (thin dot-dashed) where the transition from
stable homogeneous oscillations to phase chaos takes place. 
The curves $L_1$ (long
dashed), $L_2$ (thin dashed) and $L_3$ (dashed) as obtained in
\cite{chao1,chate} separate the various chaotic states.  The filled
circles correspond to our estimates of the $L_1$ and $L_3$ transitions
based on direct simulations of the CGLE along the 17 cuts in
coefficient space that we studied.  The open circles correspond to the
location in coefficient space where the maximal inter-peak spacing
$p_{max}$ is equal to the maximal MAW period $ P_{SN}$.  Only small
discrepancies between these two can be observed. Finally the full
curve shows the $P_{SN}\rightarrow\infty$ limit which we conjecture to be
a lower boundary for the transition from phase to defect chaos.
}\label{fig1}
\end{figure}

The main points of this paper are outlined in the following and
illustrated in Figs.~\ref{fig0},\ref{fig1}.
{\em{(i)}} Our investigation starts with the study of MAWs, which are
uniformly propagating, spatially periodic solutions of the CGLE.
These MAWs are parameterized by the average phase gradient $\nu$ 
and their spatial period $P$. 
Our study is confined to the case $\nu = 0$ for reasons specified
below. 
Spatial profiles and the stable propagation of a particular MAW are
presented in Fig.~\ref{fig0}a-c.
Isolated MAW structures consisting 
of just one spatial period $P$ play an important role in
defect formation. In particular, for fixed CGLE coefficients the range of
existence of coherent MAWs is limited by a saddle-node ($SN$) bifurcation which
occurs when $P$ reaches a maximal period $P_{SN}$. 
{\em{(ii)}} If the MAWs are driven into conditions with $P > P_{SN}$ a dynamical
instability occurs leading to the formation of defects
(Fig.~\ref{fig0}d).
{\em{(iii)}} Slowly evolving structures reminiscent of MAWs (``near-MAWs'')
are observed in the phase chaotic regime  (Fig.~\ref{fig0}e,f).
In order to characterize such states, we have examined
the distribution $D(p)$ of spacings $p$ between
neighboring peaks of the phase-gradient profile. 
In particular for sufficiently long spacing $p$,
the observed phase chaos structures are often very similar to a single 
period of a coherent MAW (Fig.~\ref{fig0}f).
{\em{(iv)}} When a phase chaotic state
displays spacings $p$ larger than $P_{SN}$, phase chaos breaks down and
defects are formed ({\it e.g.} at $t=400, x=360$ in Fig.~\ref{fig0}i). 
Thus, the MAW with $P=P_{SN}$ may be viewed as a ``critical nucleus'' for
the creation of defects. 
In phase chaos defect formation is similar to
the dynamical process by which isolated MAW structures generate defects
(Fig.~\ref{fig0}d). 
Therefore purely phase chaotic states are those for which $p$ remains 
bounded below $P_{SN}$ (Fig.~\ref{fig0}g), while defect chaos can occur when
$p$ becomes larger than $P_{SN}$ (Fig.~\ref{fig0}i). 
{\em{(v)}} A more
detailed study of the probability distribution of the $p$'s shows that for 
large $p$ the probability decays exponentially (Fig.~\ref{fig0}h,j). As long
as $P_{SN}$ has a finite value, we expect that, possibly after a very
long transient time, defects will be generated. 
{\em{(vi)}} However, in a finite
domain of the phase chaotic region, MAWs of arbitrarily large $P$
exist: we expect that in this region, even in the thermodynamic limit,
phase chaos will persist.  Fig.~\ref{fig1} shows the main quadrant of
the CGLE coefficient space.  The region of persistent phase chaos is
bounded by the Benjamin-Feir-Newell curve (thin dot-dashed) and the
curve along which $P_{SN}\to\infty$ (full curve in Fig.~\ref{fig1}). 

The outline of this paper is as follows: Section \ref{seccs} is
devoted to the study of the coherent MAW structures.  In section
\ref{sbif} we study the bifurcation diagram of the MAWs, starting from
the homogeneous oscillation. In section \ref{sdef} the incoherent
dynamics of near-MAW structures is presented. We show that for
$p\!>\!P_{SN}$, {\it i.e.,} beyond the saddle-node bifurcation,
near-MAWs evolve to defects. To illustrate the origin of the
saddle-node bifurcations in section \ref{break} we compare
bifurcation diagrams of coherent structures for different phase
gradient expansions of the CGLE.  For the lowest order expansion
(known as the Kuramoto-Sivashinsky equation \cite{kura}) the
saddle-node bifurcation is absent while it is captured by expansions
of higher order.  This explains why the divergence of the phase
gradient was exclusively observed in simulations \cite{saka} of higher
order expansions.  In section \ref{seclsc} we study various aspects of
spatiotemporal chaos in the CGLE, and relate the observed continuous
($L_1$) and discontinuous ($L_3$) transitions (see Fig.~\ref{fig1}) to
properties of the MAWs.  {\em The transition to defect chaos takes
place when near-MAWs with periods larger than $P_{SN}$ occur in a
phase chaotic state}.  In section \ref{mec} the typical values of
$p$ in the phase chaotic regime are related to the competition of two
instabilities of the MAWs, and it is possible to give a good estimate
for the numerically measured transition from phase to defect chaos
from these considerations.  A discussion of the presented results and
some final remarks are reported in section \ref{conclu}.

\section{Modulated Amplitude Waves} \label{seccs}

In this section we study the main properties of modulated amplitude
waves (MAWs) \cite{maw}. First, in section \ref{scoh} the coherent
structure framework that we use to describe the MAWs is introduced.
The bifurcation diagram of MAWs is explored in  section
\ref{sbif}, with a particular focus on the saddle-node bifurcations
that limit the range of existence of MAWs.  In section \ref{sdef}
we study the nonlinear evolution of near-MAWs that are ``pushed''
beyond their saddle-node bifurcation and show that this leads to the
formation of defects. Finally, in section \ref{break} a bifurcation
analysis of MAW-like coherent structures is performed in various {\em
phase equations} that have been proposed as approximated models for
the phase chaotic dynamics of the CGLE, and we show that only higher
order phase equations reproduce the saddle-node bifurcation.

\begin{figure} [tbph]
 \epsfxsize=.7\hsize \mbox{\hspace*{.15 \hsize} \epsffile{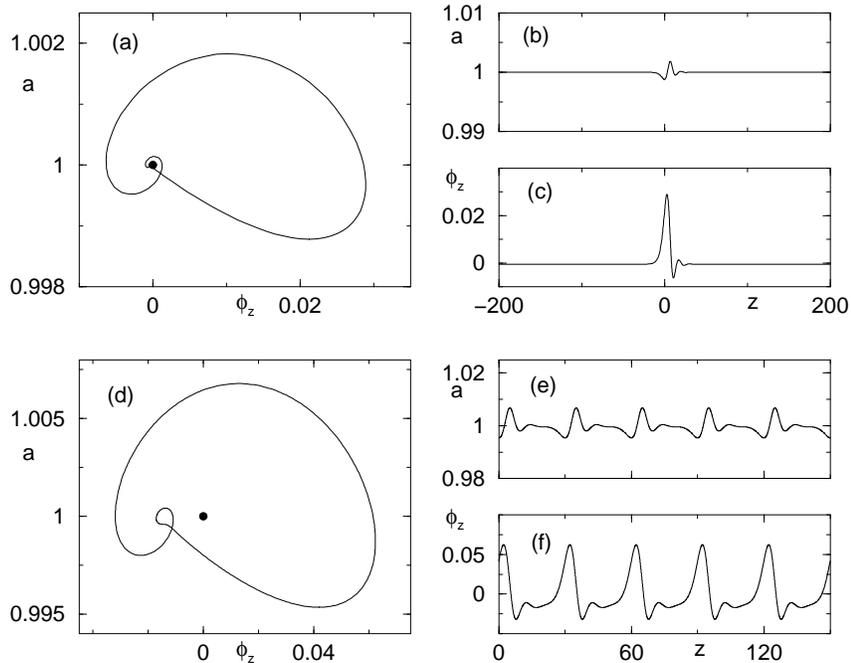} }
\vspace{-1cm} 
\caption[]{Examples of ODE solutions and corresponding
amplitude and phase gradient profiles of MAWs. (a) Homoclinic orbit
for $c_1\!=\!0.55$ and $c_3\!=\!2$; (b,c) corresponding profiles.  (d)
Limit cycle for $c_1\!=\!0.60 , c_3\!=\!2$ and $P=30$; (e,f)
corresponding profiles.  Dots in (a,d) denote the unstable fixed point
$(1,0,0)$ from which these orbits emerged.  } \label{orbits}
\end{figure}

\subsection{Coherent Structures}\label{scoh}

{\em Coherent} structures in the CGLE are uniformly propagating
structures of the form
\begin{equation}
A(x,t)= a(x-vt) e^{i \phi (x-vt)} e^{i \omega t}~,
\label{ansatz}
\end{equation} 
where $a$ and $\phi$ are real-valued functions of $z\!:=\!x - vt$.
Coherent structures have been studied extensively
\cite{torcini,saar,mvh} and play an important role in various regimes
of the CGLE \cite{pumir,miguel,torcini,saar,mvh}.

The restriction to uniformly propagating structures reduces the CGLE to a set
of three coupled ordinary differential equations (ODEs) \cite{note1}. 
These ODEs are readily found by substitution
of Ansatz~(\ref{ansatz}) into the CGLE~(\ref{cgle}) and read as:
\begin{eqnarray}
a_z &=& b \nonumber \\ 
b_z &=& \psi^2 a- \gamma^{-1} [(1-c_1 \omega)a+v(b+c_1\psi a)-(1-c_1 c_3)a^3]
\label{ode} \\
\psi_z &=& -2b\psi/a+ \gamma^{-1} [c_1+\omega+v(c_1b/a-\psi)-(c_1+c_3) a^2] ~, 
\nonumber
\end{eqnarray}
where $b\!:=\!a_z$, $\psi\!:=\!\phi_z$, and $\gamma \!:=\! 1 + c_1^2$.
Solutions
of the ODEs~(\ref{ode}) correspond to coherent structures of the CGLE.

The simplest relevant solutions of these ODEs are the fixed points
given by $ (a, b, \psi) = (\sqrt{1-q^2}, 0, q) $; these correspond to
plane wave solutions of the CGLE where $ A(x,t) = \sqrt{1-q^2} \exp
{i (q x + \omega t)} $ and $\omega = c_3-q^2(c_1+c_3)$.  An example
of more complex solutions of the ODEs (\ref{ode}) are heteroclinic
orbits which correspond to coherent structures that asymptotically
connect different states. Examples of such structures are fronts that
connect nonlinear plane waves to the homogeneous state $A\!=\!0$
\cite{saar} and Nozaki-Bekki holes that connect plane waves of
different wavenumber $q$ \cite{saar,nb}.

Here we present an extensive study of the structures that are associated 
with the {\em limit cycles} of the ODEs (\ref{ode}) \cite{notemulti}. These
limit cycles correspond to spatially periodic solutions of the CGLE
that we have already referred to as MAWs (Fig.~\ref{orbits}). For
appropriate choices of $c_1$ and $c_3$, the period $P$ of these MAWs
can be made arbitrarily large, and in this limit the limit cycles
approach a homoclinic orbit connecting the stable and unstable
manifold of one of the plane wave fixed points
(Fig.~\ref{orbits}a). Some of these infinite period MAWs have also
been referred to as ``homoclinic'' holes, and have been studied
extensively recently \cite{mvh,mm}; they are qualitatively different
from the well-known Nozaki-Bekki holes \cite{nb}.

Even if the coefficients $c_1$ and $c_3$ are fixed, MAWs are not
uniquely determined. Counting arguments, similar to those developed in
\cite{saar}, yield that in general we may expect a two-parameter
family of solutions. Let us first perform the counting for the
homoclinic orbits. As shown in \cite{mvh}, these orbits connect the
one-dimensional unstable manifold of a fixed point with its
two-dimensional stable manifold. In general, one needs to satisfy one
condition to make such a connection, in other words, such a homoclinic
orbit is of codimension one. Since the coherent structure Ansatz
(\ref{ansatz}) has two freely adjustable parameters ($\omega$ and
$v$), we therefore expect a one parameter family of homoclinic orbits.

The situation for the limit cycles of the ODEs is even simpler. Limit
cycles are of codimension zero in parameter space, and so we expect a
two parameter family of limit cycles.  In other words, if we have
found a limit cycle for certain values of $v$ and $\omega$, then we
expect this limit cycle to persist for nearby values of the parameters
$v$ and $\omega$.

Obviously, we can parameterize this family of limit cycle coherent
structures by $v$ and $\omega$, but this is not very
insightful. Instead we will use the following two quantities that are
more directly accessible in studies of the CGLE: the spatial period
$P$ of the MAWs, and their average phase gradient $\nu := (\int_0^P dx
\phi_x)/P$.  Note that for homoclinic holes, $P$ simply goes to
infinity; thus homoclinic orbits and limit cycles are members of a
single family.

The multiplicity of the MAWs can also be obtained by considering the
instability of the plane wave solutions from which the MAWs emerge
\cite{hager} (see section \ref{sbif} below).  The plane waves form a
one-parameter ($q$) family and undergo the well-known Eckhaus instability
when the coefficients $c_1, c_3$ are increased beyond certain critical
values which depend on $q$.  
In the unstable regime, a plane wave with wavenumber $q$ is
unstable to a whole band of perturbations with wavenumbers $k \in
[0,k_{max}(q)]$ \cite{KS}.  For finite systems of size $L$, this
instability thus only appears when $L > L_{min} = 2 \pi / k_{max}$.
Therefore for each $q$ there is a unique one-parameter ($L$) family of
perturbations that can render the plane wave unstable and at each of
the corresponding bifurcations a new MAW solution emerges.  Hence also
by this line of reasoning MAWs form a two-parameter family. 
%Note that
%the wavenumber $q$ of the plane wave determines the average phase
%gradient $\nu$ and the length $L$ has to be compatible with the
%spatial period $P$.
%For $L = L_{min}$, typically a MAW branch with period $P=L$ is born. 

\subsection{Bifurcation Scenario for MAWs}\label{sbif}

\begin{figure}[tbph] \vspace{0cm}
\epsfxsize=.7\hsize \mbox{\hspace*{.1 \hsize} \epsffile{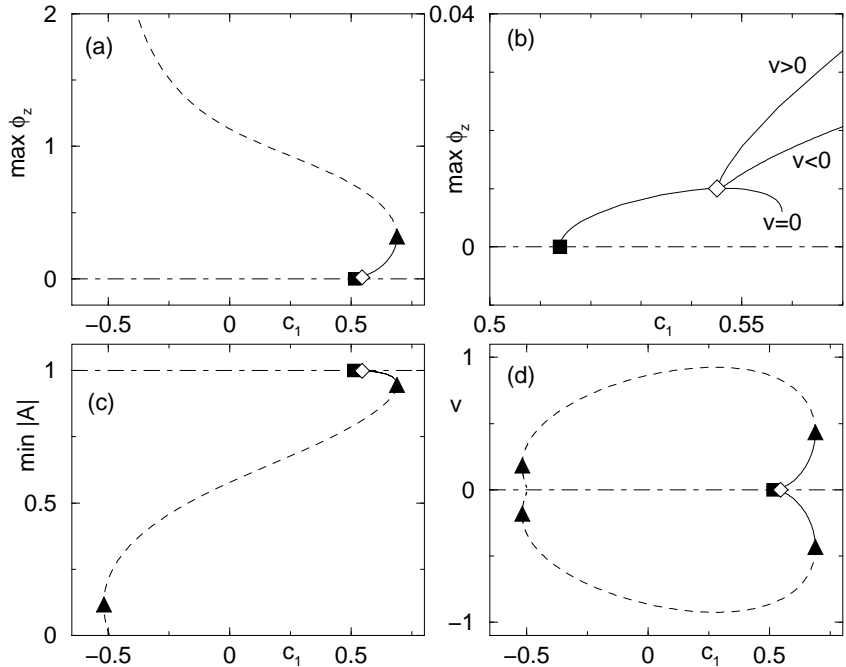}
}
\vspace{-1cm} \caption[]{Bifurcation diagrams for fixed $c_3\!=\!2$
and $P\!=\!30$, showing Hopf (filled square), drift pitchfork (open
diamond) and saddle-node (triangle) bifurcations. The dot-dashed line
represents the homogeneously oscillating solution of the CGLE, while
lower and upper branch MAWs are represented by full and dashed curves
respectively. (a) Overview of the maximum phase gradient of the MAWs
as function of $c_1$, (b) close-up, (c) the minimum of $|A|$, and (d)
the velocity $v$. For details see text.  }\label{tmpfig1}
\end{figure}

The general counting arguments given in the previous section do not
provide information on the range of existence of MAWs as a function of
the coefficients $c_1$ and $c_3$ and the parameters $\nu$ and $P$.
Here we will focus our analysis on the $\nu=0$ case since this is most
relevant for the transition to defect chaos \cite{note3}; the $\nu
\neq 0$ case will be treated elsewhere \cite{nonzero}.

All bifurcation computations have been performed with the aid of the
software package AUTO94 \cite{Auto94}.  AUTO94 can trace MAW solutions
through parameter space, and when it detects bifurcations it can
follow the newly emerging branches.  AUTO94 discretizes the ODEs
(\ref{ode}) on a periodic domain of length $L$, and $L$ will play the
role of the period $P$ of the MAWs.  Control of the average phase
gradient $\nu = \nu_0$ is implemented via the integral constraint
$\int_0^L \psi dz = L \nu_0$.  Since periodic boundary conditions
result in translational invariance, we introduce an additional
``pinning'' condition $a_z(0)=0$ in order to obtain unique solutions.

Under these conditions, the continuation procedure works as follows.
First of all, $\nu$ and $P$ are set to fixed values, and throughout
this paper we will set $\nu\!=\!0$.  Starting from a known solution
such as a plane wave or a coherent structure obtained by other means,
AUTO94 is set up to trace the MAWs along trajectories in $c_1$, $c_3$
space, while calculating the parameters $\omega$ and $v$ of these
MAWs.

\begin{figure} \vspace{0cm}
 \epsfxsize=.7\hsize \mbox{\hspace*{.15 \hsize} \epsffile{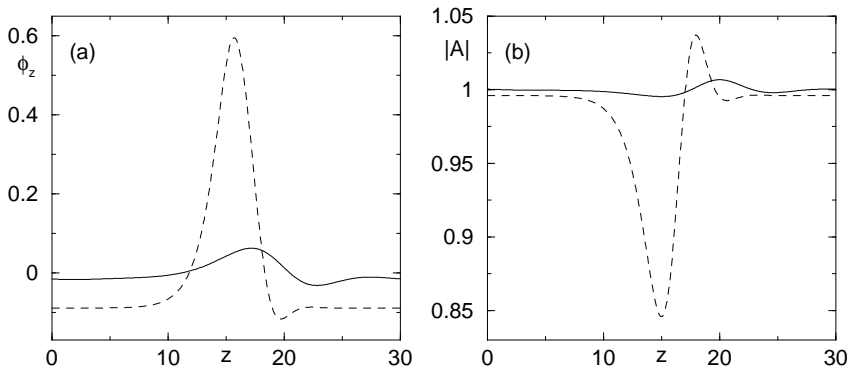} }
\vspace{-1cm}
 \caption[]{(a) Phase gradient and (b) amplitude profiles of a lower branch
 (full curve) and upper branch (dashed curve) MAW, obtained for $c_1\!=\!0.6,
 c_3\!=\!2, P\!=\!30$.  }\label{bif3}
\end{figure}

The results of our bifurcation analysis are summarized in
Fig.~\ref{tmpfig1}. When $c_1$ or $c_3$ is increased, the uniformly
oscillating state of the CGLE ($A(x,t) \!=\!  e^{i c_3 t}$) becomes
unstable via a Hopf bifurcation, from which stationary MAWs emerge
(section \ref{sshopf}).  These stationary, left-right symmetric
solutions undergo a drift pitchfork bifurcation, which leads to left
and right traveling MAWs (section \ref{ssdrift}, see also
Fig.~\ref{tmpfig1}b); as discussed later, these are the solutions
relevant for the dynamics in the phase chaotic regime.  Following
these branches of traveling MAWs, we encounter a saddle-node
bifurcation where an ``upper'' and ``lower'' branch of MAWs merge
(section \ref{sssn}, see also Fig.~\ref{bif3}); this bifurcation
limits the range of existence of MAWs and is closely related to the
formation of defects.  The upper branch MAWs can be continued back to
negative values of $c_1$, where they terminate in a solution
consisting of a periodic array of shocks and stationary Nozaki-Bekki
holes \cite{nb}. Upper branch MAWs with $P\rightarrow \infty$ have
been studied under the name homoclinic holes \cite{mvh,mm}.
 
It should be noted that, without loss of generality, we focus here on
solutions with $v\!>\!0$, for which the main peak of the phase
gradient profile is positive (see Fig.\ref{bif3}). Solutions with
$v\!<\!0$ can be obtained from right moving MAWs by applying the
mapping $x\to-x,z\to-z,v\to-v,a_z\to-a_z,\phi_z\to-\phi_z$.

\subsubsection{Benjamin-Feir instability - Hopf bifurcation}\label{sshopf}

Since the average phase gradient $\nu$ is conserved across
bifurcations, we start the continuation procedure from the uniformly
oscillating solution $A(x,t) \!=\!  e^{i c_3 t}$ that has $\nu\!=\!0$.
On an infinite domain
this uniformly oscillating solution becomes unstable via the
so-called Benjamin-Feir instability when $c_1 c_3 \ge 1$
\cite{CH}. 
In a finite domain of size $L$, the onset of this instability is
shifted to higher values of the product $c_1 c_3$
\cite{note2}; this finite size effect is relevant for our studies since 
the spatial period $L=P$ is fixed in the continuation procedure.

In the ODEs (\ref{ode}), the fixed point $(a,b,\psi)\!=\!(1,0,0)$
corresponds to the homogeneously oscillating solution.  For given
values of the period $P$, this fixed point undergoes a Hopf
bifurcation at values of $c_1$ and $c_3$ where in the CGLE
(\ref{cgle}) the mode with wavenumber $2 \pi /P$ becomes
unstable~\cite{note2}.  This Hopf bifurcation was analytically shown
to be supercritical for sufficiently small $\nu$ and large $P$ in
earlier studies \cite{janiaud,hager}; our numerical results are
consistent with this. For finite $P$, the solution bifurcating from
the fixed point is a limit cycle which approaches a homoclinic orbit
in the limit $P\rightarrow \infty$. The solutions of the CGLE that
correspond to these orbits are stationary, reflection symmetric MAWs;
an example of these is shown in Fig.~\ref{fig102}a.

\subsubsection{Drift pitchfork bifurcation}\label{ssdrift}

\begin{figure} \vspace{0cm}
 \epsfxsize=.6\hsize \mbox{\hspace*{.15 \hsize}
 \epsffile{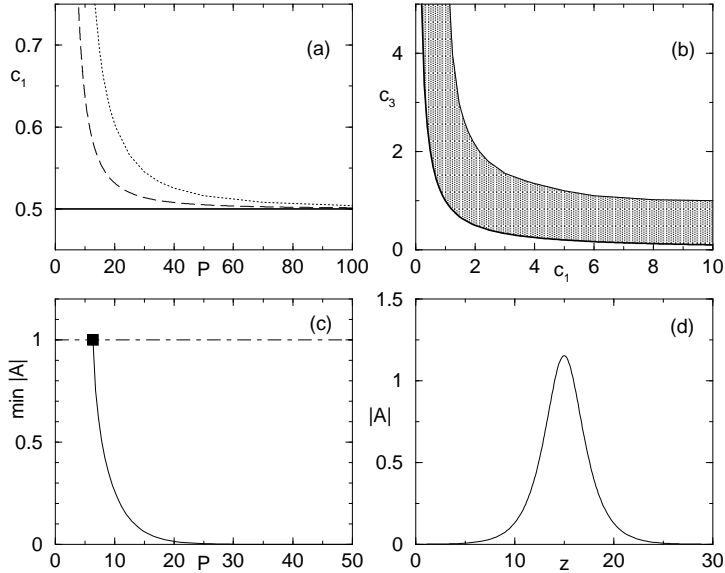} }
\vspace{-1cm}
 \caption[]{ (a) Location of Hopf (dashed curve) and drift pitchfork
 bifurcation (dotted curve) in $c_1,P$ space for $c_3\!=\!2.0$.
 (b)
 The shaded area reported in the $c_1,c_3$ coefficient space indicates
 where the drift pitchfork bifurcation does occur.
 The thick full curve in (a,b) indicates the Benjamin-Feir-Newell instability 
 for infinite domains. 
 (c) Example of a bifurcation diagram for large values of the
 coefficients $c_1=10, c_3=5$ where the drift pitchfork bifurcation does not
 occur. 
 For increasing $P$ the MAW solutions approach regular arrays of stationary
 pulses; an example of such a pulse is shown in (d) for $P=30$.
 }\label{tmpfig2}
\end{figure}

When the CGLE coefficients $c_1$ and/or $c_3$ are increased further,
the stationary MAW undergoes a drift pitchfork bifurcation
\cite{DwightDP} from which two new branches of asymmetric
($v\!\neq\!0$) MAWs emerge (see Fig.~\ref{tmpfig1}b); one of these
moves to the left, one to the right.  The locations of both the Hopf
and the drift pitchfork bifurcation approach the Benjamin-Feir-Newell
curve for large $P$ (Fig.~\ref{tmpfig2}a), while for smaller $P$ the
drift pitchfork occurs for increasingly larger coefficients $c_1$ and
$c_3$. However, only when these coefficient lie in the range shown as
the shaded area in Fig.~\ref{tmpfig2}b, the pitchfork bifurcation can
occur. Otherwise, only stationary MAWs are found. For increasing $c_1$
and $c_3$ these MAWs become pulse-like and finally approach the
solitonic solutions of the nonlinear Schr\"odinger equation
\cite{saar} (Fig.~\ref{tmpfig2}c,d).

For the case $\nu\!\neq\!0$ \cite{nonzero}, the initial plane wave
already breaks the reflection symmetry, the initial MAW has nonzero
velocity and the drift pitchfork bifurcation is replaced by its
typical unfolding \cite{ioos}.

\subsubsection{Saddle-node bifurcation}\label{sssn}

\begin{figure} \vspace{0cm}
 \epsfxsize=.7\hsize \mbox{\hspace*{.15 \hsize} \epsffile{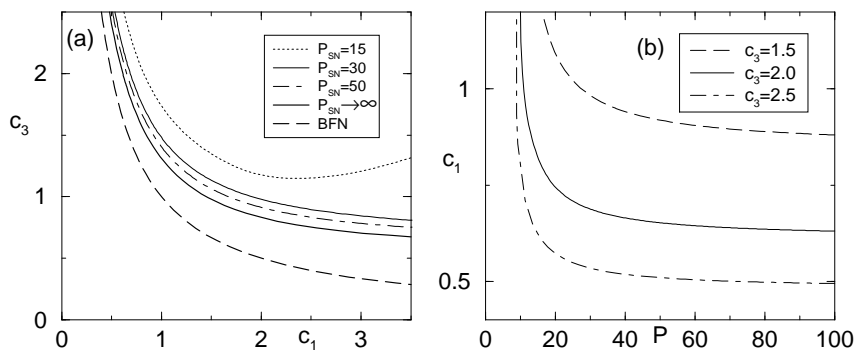} }
\vspace{-1cm}
 \caption[]{Locations of the saddle-node bifurcations in the $c_1,c_3$ plane
 (a) and the $P,c_1$ plane (b). BFN denotes the Benjamin-Feir-Newell curve.
 }\label{snhyst}
\end{figure}

Along the branch of right traveling MAWs that we described above, the
maximum of the phase gradient grows with increasing $c_1$ and $c_3$
until a saddle-node (SN) bifurcation is reached, where these MAWs
merge with another branch of MAW-like solutions.  To distinguish these
branches we refer to them as the ``lower'' and the ``upper'' branch;
for examples see Figs.~\ref{tmpfig1},\ref{fig2}.  The lower branch
MAWs are the key to understand more of the phenomenology of phase
chaos. The upper branch MAWs can, similarly to the lower branch MAWs,
be parameterized by $\nu$ and $P$, but for the same parameters, they
present more pronounced modulations (see Fig.~\ref{bif3}).

The most important aspect of the saddle-node bifurcation is that it
limits the range of existence of MAWs, since we will show that this
limit is responsible for the transition from phase to defect chaos.
Fixing $\nu\!=\!0$, the locations of these bifurcations form a
two-dimensional manifold in the three dimensional space spanned by
$c_1$, $c_3$ and $P$.  In Fig.~\ref{snhyst}a the saddle-node curves
are shown in the $c_1,c_3$ coefficient plane for a number of fixed
periods $P$; for larger $P$, the values of $c_1,c_3$ where the
bifurcation takes place decrease.  In Fig.~\ref{snhyst}b the
saddle-node curves for a number of fixed values of $c_3$ are shown in
the $P, c_1$ plane; for larger $c_3$ ($c_1$), the saddle-node occurs
for smaller values of $P$ and $c_1$ ($c_3$) \cite{note_nm}.  Once the
coefficients $c_1$ and $c_3$ are fixed, we define $P_{SN}$ as the period
for which the saddle-node bifurcation occurs. Note that there is also
a range of coefficients $c_1$ and $c_3$ (between the $P\rightarrow
\infty$ and $c_1 c_3 \!=\!1$ curve where the saddle-node bifurcation
does {\em not} occur.
 
\subsection{Evolution of perturbed MAWs}\label{sdef}

\begin{figure} \vspace{0cm}
\epsfxsize=.6\hsize \mbox{\hspace*{.15 \hsize} \epsffile{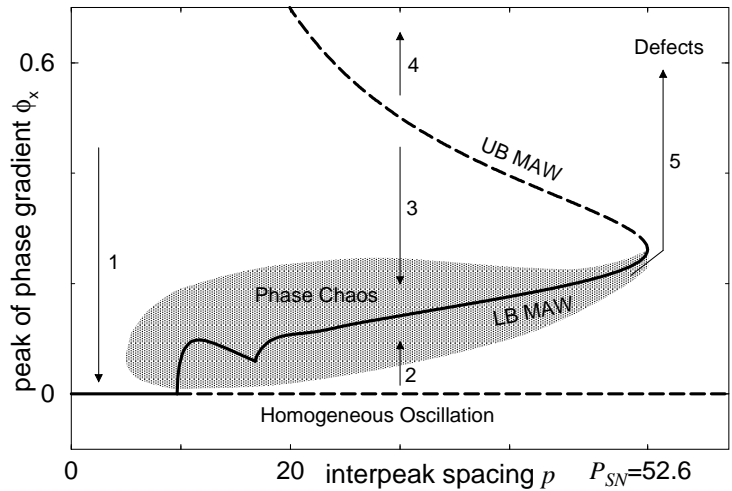}
}
\vspace{-1cm}
\caption[]{Illustration of the relation between MAW structures and
phase chaos for $c_1=0.65, c_3=2$. Both for MAWs and for an extended
profile obtained from a phase chaotic state one can extract the values
of subsequent phase gradient peaks (vertical axis) and their inter-peak
spacing (horizontal axis). The curves show the bifurcation diagram for
the lower branch (LB) MAWs and upper branch (UB) MAWs while the shaded
area indicates the typical values for near-MAW structures that occur
in phase chaos. Full (dashed) curves denote stable (unstable)
solutions for system size $L=p$. Arrows show the typical evolutions of
near-MAWs. The coefficients here are equal to those in
Fig.~\ref{fig0}~(f,i,j) and for this case we have found that phase
chaos is only a long lived transient: the shaded area reaches $P_{SN}$
just before defects appear.  }\label{fig2}
\end{figure}

In this section we will show that many basic aspects of the
phenomenology of the CGLE can be understood from a typical bifurcation
diagram of MAWs such as shown in Fig.~\ref{fig2}.  We have chosen
fixed coefficients $c_1=0.65$ and $c_3=2$ and varied the spatial
period $P$ of MAWs that exist at these coefficients.  Three families
of solutions are represented: the homogeneous oscillation, the lower
branch (LB) and the upper branch (UB) MAWs.  The shaded area
schematically indicates the near-MAW structures observed in phase
chaotic states such as shown in Fig.~\ref{fig0}~(f,i,j).  The arrows
in Fig.~\ref{fig2} represent the dynamical evolution of perturbed
MAWs, and their direction can be obtained by performing a linear
stability analysis.

{\em Linear stability - } As discussed in section \ref{sbif}, the
homogeneous solution is stable against short wavelength perturbations
(arrow 1), and turns unstable via the Hopf bifurcation that also
generates the lower branch MAWs (arrows 2). As discussed in
\cite{mvh,mm}, upper branch MAWs have at least one unstable
eigenvalue, and the dynamical evolution of perturbations is directed
away from upper branch MAWs (arrows 3,4).

\begin{figure} \vspace{-.0cm}
  \epsfxsize=.7\hsize \mbox{\hspace*{.15 \hsize} \epsffile{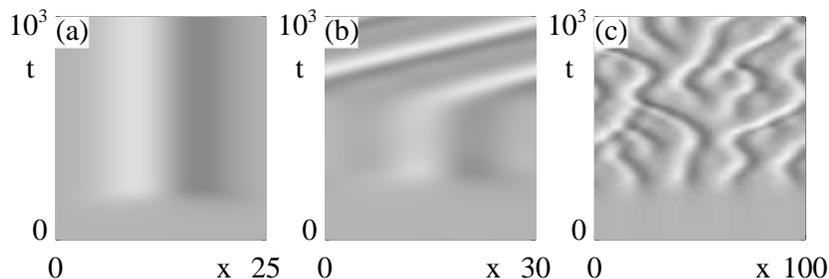}
    } \vspace{-5cm}
\caption[]{The evolution of an unstable homogeneous state towards
lower branch MAW dynamics, for $c_1\!=\!3$ and $c_3\!=\!0.6$. The
coefficients $c_1$ and $c_3$ are chosen such that no saddle-node
bifurcation occurs for any $P$.  (a) Evolution towards a stable
stationary lower branch MAW for system size $L\!=\!25$ and (b) towards
a stable drifting lower branch MAW for system size $L\!=\!30$. Note
that for the coefficients chosen, the drift pitchfork bifurcation
occurs at $P=27.7$.  (c) Evolution towards phase chaos for system size
$L\!=\!100$. Incoherent evolution of structures characterized by local
concentrations of phase gradients can be clearly observed.  We think
of these structures as ``near'' MAWs.}\label{fig102}
\end{figure}

The linear stability of lower branch MAWs will be discussed in more
detail in section \ref{mec}. It turns out that perturbations of lower
branch MAWs can evolve in many ways, but in almost all cases the
ensuing dynamics remains close to the lower MAW branch (shaded area in
Fig.~\ref{fig2}).  The only exception we have found to this rule is
when a MAW is pushed beyond the saddle-node bifurcation (arrow 5).

{\em Nonlinear evolution - }
Here we want to go beyond the linear analysis and study the nonlinear
evolution of MAWs along the arrows of Fig.~\ref{fig2}. 
The examples (at different choices of the coefficients) 
of the dynamics shown below are not exhaustive, but
should serve to illustrate typical behavior which 
appears to be very robust.

{\em arrow 2 - } When the uniform oscillation becomes linearly
unstable perturbations grow.  To the left of the saddle-node,
perturbations evolve to dynamics dominated by lower branch MAWs
(Fig.~\ref{fig102}). For small system sizes, stable MAWs may occur
(Fig.~\ref{fig102}a,b), while for larger systems periodic sequences of
MAWs are unstable with respect to the so-called {\em interaction} or
{\em splitting} instabilities \cite{maw,chang} that will be discussed
in section \ref{mec}. Hence a perturbed unstable homogeneous state
typically does not converge to a train of coherent MAWs, but instead
evolves to phase chaos (Fig.~\ref{fig102}c).  In the context of the
bifurcation diagram, note that the disordered structures observed in
the phase chaotic evolution are quite similar to lower branch
MAWs. The shaded area in Fig.~\ref{fig2} represents this ``near-MAW''
behavior.

\begin{figure} \vspace{-.20cm}
  \epsfxsize=.7\hsize \mbox{\hspace*{.15 \hsize}
    \epsffile{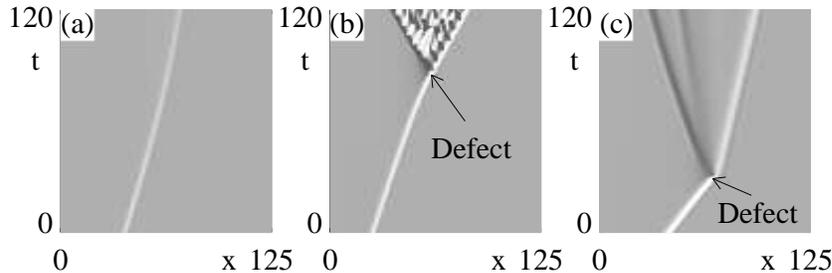} } \vspace{-5.cm}
\caption[]{Evolution of perturbations of the upper branch MAWs. The
coefficients $c_1$ and $c_3$ are chosen such that no saddle-node
bifurcation occurs for any value of $P$. (a) A slowing down and
spreading of the phase gradient characterizes the decay to a lower
branch MAW for $c_1\!=\!0.55$ and $c_3\!=\!2$. (b) For the same
coefficients, another perturbation leads to an increase in velocity
and divergence of the phase gradient. A defect occurs, from which
hole-defect dynamics spreads for these coefficients. (c) For
$c_1\!=\!3$ and $c_3\!=\!0.6$ a perturbed upper branch MAW leads to a
defect, but defects do not percolate through the
system. }\label{fig106}
\end{figure}

{\em arrows 3,4 - } Upper branch MAWs are always unstable due to the
positive eigenvalue associated with the saddle-node bifurcation. The
resulting incoherent dynamics has been studied quite extensively in
the context of hole-defect dynamics \cite{mvh,mm}. {\em{(i)}} When a
perturbation has pushed an upper branch MAW towards the ``lower'' part
of the bifurcation diagram, the structure decays towards lower
branch MAWs (arrow 3). An example of a
space time plot for the decay towards a lower branch MAW is shown in
Fig.~\ref{fig106}a. {\em{(ii)}} When the perturbation pushes the MAW
towards the ``upper'' side of the diagram, the phase gradient peak that
characterizes MAWs grows without bound, and at the same time the
minimum of $|A|$ approaches zero: a defect is formed (arrow 4). The
dynamics {\em after} such a defect has formed depends on the values of
the coefficients $c_1$ and $c_3$. Two different examples are shown in
Fig.~\ref{fig106}b,c. For more details see section \ref{seclsc}.

{\em arrow 5 - } So far we have encountered two scenarios: if the
phase gradient peak of a structure is ``larger'' than that of an upper
branch MAW, then it will grow out to form defects. If it is
``smaller'', it will decay back in the direction of the lower branch
MAWs. The latter process frequently occurs in phase chaos, preventing
the formation of defects, while the former process needs to be
initiated by appropriate initial conditions. However, when the upper
and lower branches approach each other and disappear in a saddle-node
bifurcation, there are no structures left to prevent arbitrary small
perturbations to grow out to defects. This dynamical process, which is
represented by arrow 5 in Fig.~\ref{fig2}, is the core of our
argument: defect formation takes place beyond the saddle-node
bifurcation.

\subsection{Breakdown of phase description}\label{break}

An alternative approach to describe the creation of defects from phase
 chaotic states is via blow-ups in so-called phase-equations
 \cite{saka}. Phase equations are based on the observation that close
 to the onset of phase chaos (near the Benjamin-Feir-Newell curve) the
 amplitude is ``slaved'' to the phase dynamics. In this situation a
 phase equation can be obtained by a gradient expansion \cite{KS}.
 The expansion including all parity-symmetric terms up to fourth order
 \cite{saka} reads
\begin{eqnarray}
\frac{\partial\phi}{\partial t}&=&\Omega_2^{(1)}\frac{\partial^2\phi}{\partial x^2}
   + \Omega_2^{(2)}\left(\frac{\partial \phi}{\partial x}\right)^2
   + \Omega_4^{(1)}\frac{\partial^4\phi}{\partial x^4} 
\label{sakaequ} \\
& &+ \Omega_4^{(2)}\frac{\partial\phi}{\partial x}
     \frac{\partial^3\phi}{\partial x^3} +
     \Omega_4^{(3)}\left(\frac{\partial^2\phi}{\partial x^2}\right)^2 +
     \Omega_4^{(4)}\left(\frac{\partial\phi}{\partial x}\right)^2
     \frac{\partial^2\phi}{\partial x^2} \nonumber
\end{eqnarray}
where $\Omega_2^{(1)}\!\!=\!\!1-c_1 c_3,
\Omega_2^{(2)}\!\!=\!\!-(c_1+c_3), \Omega_4^{(1)}\!\!=\!\!-c_1^2
(1+c_3^2)/2, \Omega_4^{(2)}\!\!=\!\!-2 c_1 (1+c_3^2),
\Omega_4^{(3)}\!\!=\!\!-c_1 (1+c_3^2), \Omega_4^{(4)}\!\!=\!\!-2
(1+c_3^2)$. The lowest order description of phase chaos is obtained
when the parameters $ \Omega_4^{(2)}, \Omega_4^{(3)}$ and $
\Omega_4^{(4)}$ are set equal to zero; the resulting equation is known
as the Kuramoto-Sivashinsky equation \cite{kura}.

The phase equations with higher order terms included have been studied
via direct integration by Sakaguchi~\cite{saka}. For the full
Eq. (\ref{sakaequ}), Sakaguchi observed finite time divergences of the
phase gradient for coefficients close to the transition from phase to
defect chaos in the CGLE.  He attributed such divergences to the
occurrence of defects in the CGLE. No blow-up of the phase gradient is
observed for Eq.  (\ref{sakaequ}) without the last term, or for the
simple Kuramoto-Sivashinsky equation. Recently, Abel {\it et al.}
\cite{abel} quantified the increasing discrepancies between the phase
equations of different orders and the full dynamics in the CGLE with
increasing distance from the Benjamin-Feir-Newell curve and identified
the relative importance of the various terms in Eq. (\ref{sakaequ}).

\begin{figure} \vspace{0cm}
\epsfxsize=.7\hsize \mbox{\hspace*{.15 \hsize} \epsffile{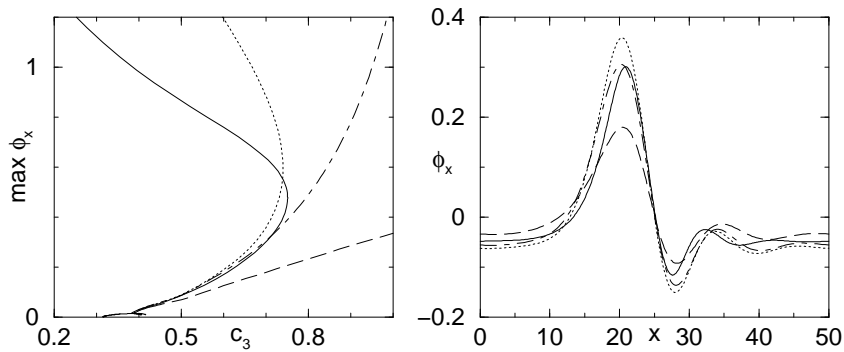} }
\vspace{-1cm}
\caption[]{ Comparison of different phase expansions :
Eq.(\ref{sakaequ}) (dotted), Eq.(\ref{sakaequ}) without the last term
(dot-dashed), Kuramoto-Sivashinsky Eq. (dashed) and CGLE (full curve).
Parameters are $c_1=3.5, \nu=0, P=50$.  (a) Bifurcation diagrams and
(b) spatial profiles of lower branch coherent structures at $c_3=0.7$.
}
\label{phedp}
\end{figure}

Since the essential ingredient of our theory is the occurrence of a
saddle-node bifurcation, we have investigated the bifurcation scenario
for various truncations of the phase equations (\ref{sakaequ}). In the
context of phase dynamics, our Ansatz (\ref{ansatz}) becomes of the form
\begin{equation}
\phi(x,t)=\tilde{\phi}(x-v t)+(\omega-c_3) t~.
\end{equation}
We have studied MAW-like structures occurring in the phase equations by
employing the same methodology as for the CGLE; the average phase
gradient value $\nu$ is fixed to $0$ and $P$ parameterizes the spatial
period of the MAW.  In Fig.~\ref{phedp} we compare bifurcation
diagrams and MAW profiles for different expansions at the parameters
$c_1=3.5, P=50$.

For all phase equations considered here the coherent structures are
again born in a Hopf and undergo a drift pitchfork bifurcation, beyond
which the maximal phase gradients increase.  This leads to increasing
discrepancies between different approximations.  In particular, the
coherent structures for Eq. (\ref{sakaequ}) exhibit saddle-node
bifurcations at parameter values not far from those for corresponding
MAWs in the CGLE; nevertheless the MAWs of Eq. (\ref{sakaequ}) deviate
substantially from the CGLE MAWs for the upper branch of MAWs.  The
Kuramoto-Sivashinsky equation, and Eq. (\ref{sakaequ}) without the
last term, do not exhibit a saddle-node bifurcation.  Since these
latter two models do not experience blow-up, we can safely conclude
that these observations confirm our picture, and that the saddle-node
bifurcations of coherent structures play the same crucial role in both
the full CGLE and its phase equations.

\section{Large scale chaos} \label{seclsc}

In this section we will study the dynamical evolution of the CGLE near
the transitions from phase to defect chaos.  The transition between
these two states can either be hysteretic or continuous: in the former
case, the transition is referred to as $L_3$, in the latter as $L_1$
\cite{note4}.

\begin{figure} \vspace{-.20cm}
  \epsfxsize=1.\hsize \mbox{\hspace*{.15 \hsize} \epsffile{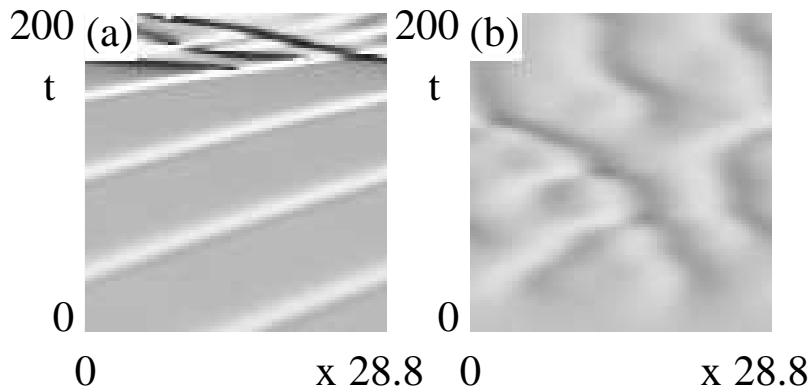}
    } \vspace{-7cm}
\caption[]{ Defect formation at $c_3=2, c_1=0.7$.  (a) Defect
formation. As initial condition we took a lower branch MAW with
$P<P_{SN}\!=\!26.8$ which we embedded in a background of zero
wavenumber. The system size $L$ here is equal to 28.8, which is larger
than $P_{SN}$ and a defect is formed; for $L<P_{SN}$ this defect
formation does not take place. (b) Random initial conditions in
general evolve to MAW like structures with $P<P_{SN}$ which do not
lead to defects; the ``critical'' nucleus that leads to defect
formation has a rather small basin of attraction here.
}\label{fig107}
\end{figure}

How are defects generated from phase chaos? Let us start to consider a
{\em small} system in which a stable lower branch MAWs has been
created. When we fix the coefficients $c_1$ and $c_3$ and steadily
increase the size of the system, and hence the period $P$ of the MAW,
we find that as soon as we push $P$ beyond $P_{SN}$, the MAW structure
blows up to form defects. An example of this is shown in
Fig.~\ref{fig107}a. In a similar fashion, defects are created when the
system size $L$ is fixed, and either $c_1$ or $c_3$ are increased
until $P_{SN}\!<\!L$ (Fig.~\ref{fig0}c,d).

How is this related to phase chaos? As shown in Fig.~\ref{fig107}b,
typical phase chaotic states show much more incoherent dynamics,
containing many MAW like structures but of much smaller period. Our
central conjecture is therefore that the transitions from phase to
defect chaos are triggered by the occurrence of near-MAW structures in
a phase chaotic state with $\nu\!=\!0$ \cite{note3} and periods larger
than $P_{SN}$, the spatial period of the critical nucleus for defect
creation.

To test this conjecture, we have numerically investigated the
distribution of inter-peak spacings $p$ of the phase gradient profile
(see Fig.~\ref{fig0}e,f). In section \ref{ss3.1} we discuss the
definition of $p$ and the details of our numerical analysis.  In
particular, we have examined in the $c_1,c_3$ plane 17 different
``cuts'' across the $L_1$ and $L_3$ transition lines.  In section
\ref{ssL1} the results of our numerics along a cut through the $L_1$
transition line are presented, while section \ref{ssL3} is devoted to
the $L_3$ transition.  We will show that the presence of inter-peak
spacings $p$ larger than $P_{SN}$ accurately predicts the transition
from phase to defect chaos (Fig.~\ref{fig1}).  In the last section
\ref{mec} we will show that a reasonable, parameter-free estimate of
the numerically observed transitions can be obtained via a linear
stability analysis of the MAWs.

\subsection{Identification of MAWs in the phase-chaotic regime}\label{ss3.1}

\begin{figure} \vspace{0cm}
\epsfxsize=.7\hsize \mbox{\hspace*{.15 \hsize}
\epsffile{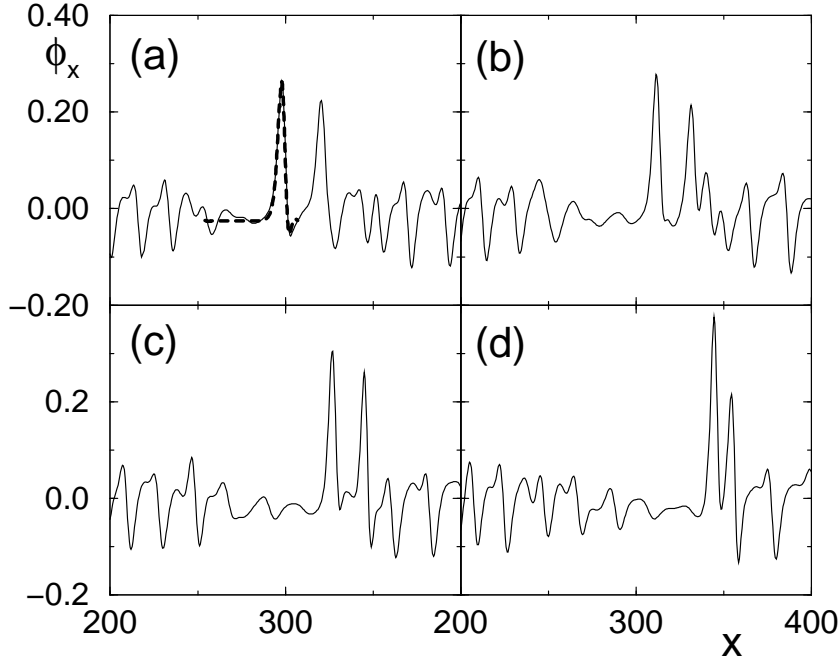}}
\vspace{-1cm}
\caption[]{Local phase gradient of the chaotic field just before
defect formation for $c_1=0.65$ and $c_3=2$. Panel (a) is a snapshot
of the field at a time $t=120$ before the occurrence of the defect,
(b) , (c) and (d) are successive snapshots taken at time intervals
$\delta t = 30$. In (a) also the shape of the MAW at the saddle-node
is superimposed (thick dashed line) on the profile.}
\label{defcrea.2b0}
\end{figure}

To verify our main conjecture, we have to characterize the MAW
structures occurring in the phase-chaotic regime.  In general this is
a complicated task, since the phase gradient profile of a typical
phase chaotic state (see Fig.~\ref{fig0}e,f and \ref{defcrea.2b0})
consists of many peaks of different size, spacing and shape; a priori
it is unclear how to compare these to MAW profiles.  However, a close
inspection of the defect forming process reveals that while closely
spaced phase gradient peaks evolve in a quite erratic way, well spaced
peaks appear to have a more regular dynamics and frequently their
overall shape resembles that of MAWs (see Fig.~\ref{defcrea.2b0}).
These large period near-MAWs modify their shape quite slowly with
respect to the other structures present in the chaotic field, and
propagate over a disordered background. Therefore we
study the distribution of inter-peak distances $p$, keeping in mind
that the tail of this distribution is relevant for defect generation.

The phase gradient profile of a coherent MAW (see Figs.~\ref{fig0}a
and Fig.~\ref{bif3}a) shows a secondary maximum.  To obtain the
correct period $P$ of a near-MAW, such small extrema should be
neglected when the inter-peak spacing $p$ is measured.  We introduce a
cutoff for the size of the phase gradient peak equal to the size of
the secondary extremum of the MAW with the largest $P$. As an
additional result of this cutoff, small fluctuations are not
considered as MAW peaks. It should be noted that the tail of the
distribution of $p$ is rather insensitive to the precise value of this
cutoff.

In order to estimate the probability density $D(p)$, for every time
interval $\tau=0.5$, the inter-peak periods $p$ of the spatial profile
of the phase gradient are determined. In addition, for every snapshot
the largest value $p_{max}$ of the inter-peak spacing $p$ is stored
separately, and this leads to the distribution $D(p_{max})$. From the
spatial profile of $|A|$ the distribution $D(|A|)$ and the minimal
amplitude value $|A|_{min}$ can be derived. This latter quantity is
used to detect defects: when $|A|_{min}$ falls below a value of 0.1,
we take this as an indication of a defect.

Extensive simulations have been made possible thanks to an innovative
time-splitting code which ensures precision and stability comparable
with pseudo-spectral codes, but is noticeably faster \cite{torcini}.
The spatial resolution $\Delta x$ has been set to $0.5$ and the
integration time step to $0.05$.  Simulations have been carried out
for integration times ranging from $t=5\times 10^5$ to $t=3 \times
10^7$ and for a typical system size $L=512$; occasionally, runs have
been performed with $L=100, 200$ and $5000$.  Typically, our runs
start from random initial conditions of the type $A_k(t=0)=|A|_k(t=0)
\cdot {\rm e}^{i \phi_k(t=0)}$ (where $A_k(t)=A(k\Delta x,t)$ and
$\phi_k(t)=\phi(k \Delta x ,t)$) with
\begin{eqnarray}
|A|_k(0) =  1 + r_k    
  ~ \label{ic}\\
\phi_k(0) = \phi_{k-1}(0)*0.8+q_k
  ~ \label{ic2}
\end{eqnarray}       
where $r_k$ and $q_k$ are random numbers uniformly distributed in
$[-0.05,+0.05]$ and $\phi_1(0)=0.0005$. This initial condition
(\ref{ic2}) leads to a smooth phase and the formation of defects due
to initial discontinuities is avoided.

In sections \ref{ssL1} and \ref{ssL3} we will consider in detail two
particular cuts in the $(c_1,c_3)$ coefficient space, one across the
$L_1$ and one across the $L_3$ curve. In particular, we will analyze
the behavior of the probability densities $D(|A|)$, $D(p)$ and
$D(p_{max})$ for both transitions.

\subsection{$L_1$ transition}\label{ssL1}

In this section we concentrate on the $L_1$ transition that is
observed when the value of $c_1$ is fixed at 3.0 and $c_3$ is varied.

{\em Transition to defect chaos - } Starting from random initial
conditions we have integrated the dynamics of the CGLE for long
durations.  For a fixed system size $L$ we observe that, as a function
of the total integration time, the value of $c_3^*$ for which defects
are formed appears to decrease.  Similar behavior occurs when the
system size $L$ is increased for fixed integration times.  For
example, for an integration time of $3 \times 10^7$ and $c_1=3$ we
find for system size $100$, $200$ and $512$ critical values $0.82,
0.81$ and $0.79$, respectively. For a size $L=5000$ and integration
times $3 \times 10^6$ a critical value of $0.79$ is also found.

Note that even the lowest value of $c_3^*$ for the numerically
measured transition obtained here is far above the lower bound
$c_3^\infty = 0.704$ which is the value of $c_3$ where the size of the
critical nucleus for defect formation diverges ($P_{SN} \to \infty$).
Below, we will give an estimate of the critical value ${\hat c}_3$ for
which the defect density should vanish in the thermodynamic limit by
extrapolating finite time and finite size data.

\begin{figure} \vspace{0cm}
\epsfxsize=.7\hsize \mbox{\hspace*{.15 \hsize}
\epsffile{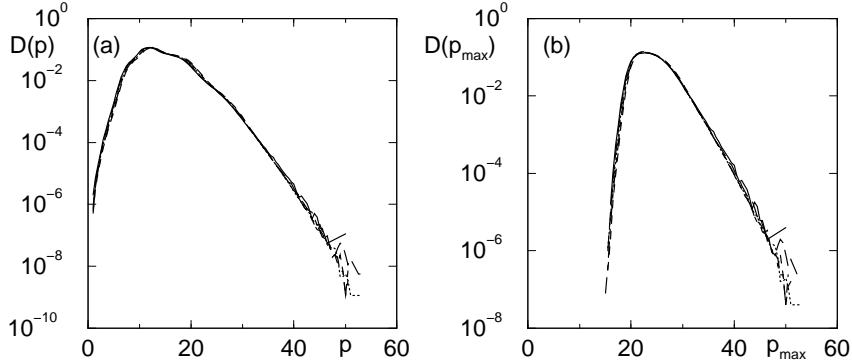}}
\vspace{-1cm}
\caption[]{ Probability densities (a) $D(p)$ and (b) $D(p_{max})$ for
$c_1=3$ and various values of $c_3$ on a lin-log scale. The curves
refer to $c_3$ below $c_3^*=0.79$ (namely to $c_3 = 0.77$ and $0.78$),
as well as to values corresponding to the defect chaotic regime: $c_3
= 0.79$, $0.80$ and $0.81$.  The system size was $L=512$ and the
integration times where $t=5\times 10^5$ for $c_3=0.81$, $t=5\times
10^6$ for $c_3=0.80$ and $t=25\times 10^6$ for all other values.  }
\label{distL1}
\end{figure}

{\em Distribution of $p$ - } Let us now consider the distribution of
$p$'s for various coefficients $c_3$ near the $L_1$ transition. It is
clear from the data reported in Fig.~\ref{distL1} that the shape of
these distributions is quite insensitive to the presence or absence of
defects. This can be partly explained by the fact that just above the
$L_1$ transition defects arise in the system as rare isolated events
occurring during the spatio-temporal evolution, as shown in
Fig.~\ref{fig106}c.  This is fully consistent with earlier
observations that the $L_1$ transition is continuous
\cite{chao1,chate,egolf}.  We focus on the tail of the probability
density $D(p)$, since this gives information on the probability to
observe defects. Our numerical results suggest an exponential decay,
{\it i.e.,} $D(p) \propto \exp(-\alpha \cdot p)$ with $\alpha=0.6$ for
sufficiently large $p$.

Similarly to the apparent transition value $c_3^*$, the values
associated to extremal events $|A|_{min}$ and $p_{max}$ depend on
integration times and system sizes.  By assuming that $D(p)$ remains
finite (but likely exponentially small) for large $p$, we can expect
that for long enough times, rare events associated with large values
$p$ will occur, and hence, defects can form after possibly
very long transients.

{\em Crossover behavior - } A good order parameter to identify the
occurrence of the transition starting from the defect chaos phase near
the $L_1$ transition is the {\em defect density} $\delta_D$ which
measures the number of defects occurring per space and time unity. In
the defect chaos regime $\delta_D >0$, while it vanishes at the
$L_1$-transition. Now we can relate this order parameter to the tail
of the distribution of $p$. Our conjecture states that defects should
arise when $p > P_{SN}$, therefore the defect density $\delta_D$
should be related to the probability to have structures of period $p >
P_{SN}$, {\it i.e.},
\begin{equation}
\delta_d \propto \int_{P_{SN}}^\infty dp D(p) \propto {\rm e}^{-\alpha
P_{SN}}~ ;
\label{defect}
\end{equation}   
where $D(p) \propto \exp(-\alpha \cdot p)$ has been used. If we now
assume that the distribution $D(p)$ does not vary significantly across
the transition (as is evident from Fig.~\ref{distL1}), then the change
in the probability to have $p>P_{SN}$ is dominated by the changes in
$P_{SN}$ with $c_3$. A reasonable fit of our bifurcation data for
$P_{SN}$ (see Fig.~\ref{snhyst}) in the interval $30 \le P_{SN} < 300$
is
\begin{equation}
   P_{SN} \approx \frac{\beta}{c_3-c_3^\infty} ~, \label{psn_c3}
\end{equation}       
where $\beta \approx 4.38$. Combining this result with the Ansatz
(\ref{defect}), we immediately obtain the following expression for the
defect density:
\begin{equation}
\delta_d \propto {\rm e}^{- \alpha \beta/(c_3- c_3^\infty)} ~.
\label{defect3}
\end{equation}       
A similar expression was proposed in \cite{chao1,egolf} for the defect
density near the $L_1$ transition.

In order to verify if the expression (\ref{defect3}) is reasonable
also for our choice of the parameters, we have estimated the
probability~\cite{torcini}
\begin{equation}
  w(|\hat{A}|) = \int_0^{|\hat{A}|} d|A| D(|A|)~, \label{defdens}
\end{equation}
to observe an amplitude less than $|\hat{A}|$.  This quantity gives a
more precise characterization of the $L_1$-transition than $\delta_D$,
because it measures not only the extreme events corresponding to true
defects, but also the tendency of the system to generate structures
characterized by small $|A|_{min}$.  We estimated the quantity
(\ref{defdens}) for several $|\hat{A}|$ values and for various $c_3$
parameter values in the defect chaos regime.  Reporting $\ln
[w(|\hat{A}|)]$ as a function of $1/(c_3-{\hat c}_3)$ a reasonable
linear scaling is observed in the range $0.795 \le c_3 \le 0.85$, for
$0.1 \le |\hat{A}| \le 0.5$, with the choice ${\hat c}_3=0.72$.  The
value ${\hat c}_3$ where the defect density should asymptotically
vanish is much smaller than $c_3^*$ obtained via direct numerical
simulations but still bigger than $c_3^\infty = 0.704$ where
$P_{SN}\to\infty$. 

We can now easily estimate the integration time needed to observe a
tiny shift of the apparent value $c_3^*$ towards the corresponding
asymptotic value $c_3^\infty \approx 0.704$. Limiting our analysis to
system size $L=512$, a typical time-scale to observe a defect at
$c_3\!=\!0.79$ is $t \sim 3 \times 10^{7}$.  At this value of $c_3$,
$P_{SN}\!=\!46.5$, while for $c_3\!=\!0.739$, $P_{SN}\!=\!105$ .
Invoking the exponential decay of $D(p)$, one immediately finds that
the time scale to observe a defect at $c_3\!=\!0.739$ is of order
$10^{17}$, which is completely outside the reach of present day
computers.

\begin{figure} \vspace{0cm}
\epsfxsize=.7\hsize \mbox{\hspace*{.15 \hsize}
\epsffile{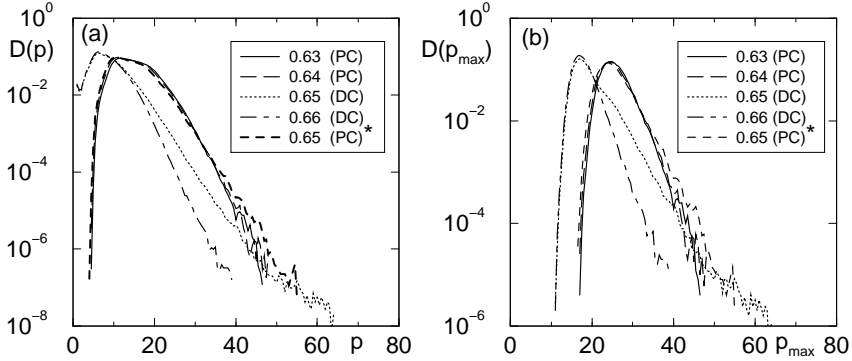}}
\vspace{-1cm}
\caption[]{ Probability densities (a) $D(p)$ and (b) $D(p_{max})$ for
$c_3=2$ and various values of $c_1$ reported in a lin-log scale.  The
data are for a system size $L=512$ and for integration times ranging
from $t=5\times 10^5$ for $c_1=0.63$,0.64,0.66 and 0.65 (PC)$^\star$
to $t=2.5 \times 10^6$ for $c_1=0.65$ (DC).  The labels DC and PC
indicate that we are in presence or absence of defects,
respectively. The label (PC)$^\star$ refers to the regime before
defect formation at $c_1=0.65$.  }
\label{distL3}
\end{figure}                      

\subsection{$L_3$ transition}\label{ssL3}

In order to characterize the $L_3$ transition from phase to defect
chaos in more detail $c_3\!=\!2$ has been fixed, while the coefficient
$c_1$ is varied.  The $L_3$ transition is hysteretic
\cite{chao1,chate}: to the left of $L_3$ one may have phase or defect
chaos depending on the initial conditions. Beyond the $L_3$ phase
chaos breaks down and defects occur spontaneously for any initial
condition.  In order to study the dynamics across this transition we
therefore initialized the simulations with initial conditions
(\ref{ic}),(\ref{ic2}) or used relaxed phase chaos configurations
corresponding to values of $c_1$ far below the $L_3$ line.

The probability densities $D(p)$ and $D(p_{max})$ are shown in
Fig.~\ref{distL3}.  For $c_1 \!<\! c_1^*=0.65$ all distributions
collapse on a unique curve, but as soon as defects arise the
distributions change substantially.  Whenever a defect is generated,
hole-defect dynamics takes place (see Fig.~\ref{fig106}b).  As a
result phase chaos is replaced by defect chaos.  The noticeable
modification of the distributions thus reflects the fact that the
$L_3$ transition is discontinuous.  Also the probability density for
$|A|$ changes abruptly across the $L_3$ transition.

\begin{figure}
\epsfxsize=1.0\hsize \mbox{\hspace*{.1 \hsize}
\epsffile{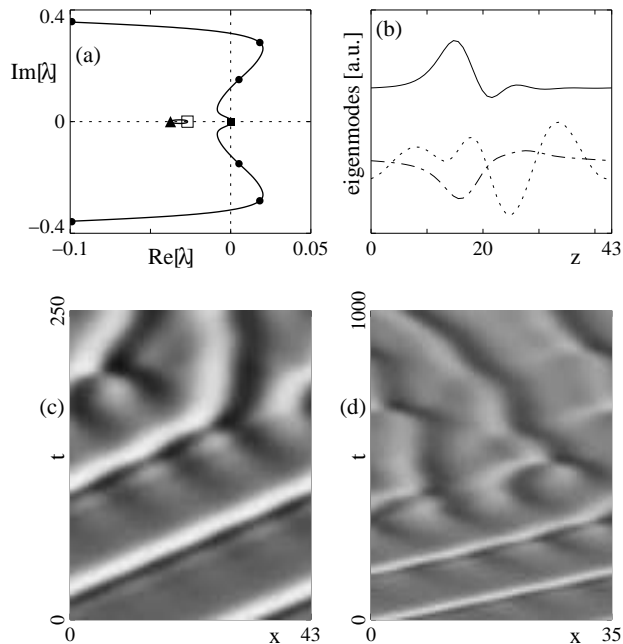} } \vspace{-4.0cm} \caption[]{Results of the
linear stability analysis : (a) leading part of the eigenvalue
spectrum (continuous spectrum denoted by the full curve, Goldstone
modes by a filled square, saddle-node by filled triangle, interaction
by open square and splitting modes in $L=P$ by dots, respectively),
(b) splitting eigenmodes (dot-dashed and dotted,
$\lambda_{split}=0.018\pm0.28i$) of the phase in $L=P$ compared with
spatial MAW profile of the phase gradient (full curve).  (c,d) Space
time plots showing the splitting of a MAW initially perturbed by small
noise.  Parameters are $c_1=3, c_3=0.72, P=43$ near $L_1$ for (a-c)
and $c_1=0.65, c_3=2,P=35$ near $L_3$ for (d).  }\label{specspl}
\end{figure}                                              

\subsection{Mechanism for the selection of $p$}\label{mec}

When approaching the transition to defect chaos from the
Benjamin-Feir-Newell curve, three parameter regions, corresponding to
different dynamical regimes, can be distinguished
(Fig.~\ref{bounds}). The first encountered region corresponds to
infinite values of $P_{SN}$ : here we expect no defects to occur,
irrespectively of system size and integration time. The phase chaos is
the asymptotic regime in this first region.  Then, when $c_1$ and/or
$c_3$ are increased, a crossover regime is reached where extreme
events (large inter-peak spacings) may lead to defect formation. Here
phase chaos can persist as a long lived transient, but eventually we
expect it to break down. Then, when $c_1$ and/or $c_3$ are even
further increased, we experience a dramatic drop in transient times,
and defect chaos sets in quite rapidly. We understand this drop to
occur when typical values of $p$ (and not rare extreme events) become
larger than the corresponding $P_{SN}$ values.

\begin{figure}
\epsfxsize=1.0\hsize \mbox{\hspace*{.1 \hsize}
\epsffile{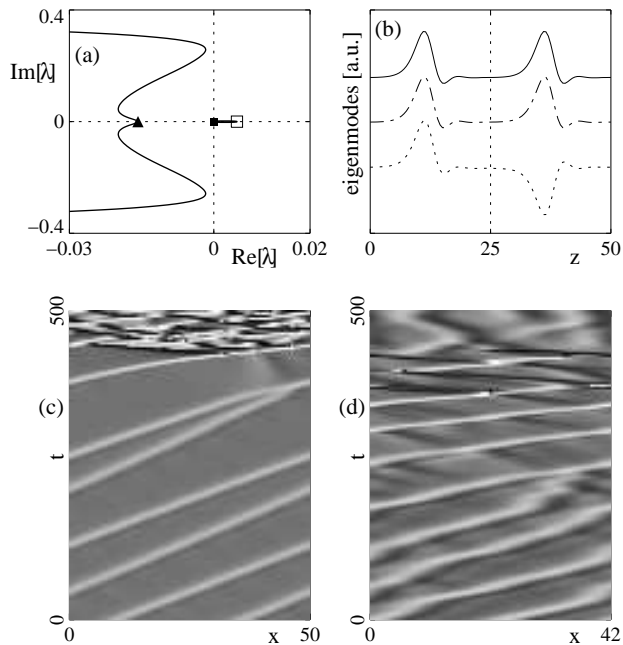} } \vspace{-4.0cm}
\caption[]{Results of the linear stability analysis :
(a) leading part of the eigenvalue spectrum 
(continuous spectrum denoted by the full curve,
Goldstone modes by a filled square,
saddle-node by filled triangle and interaction by open square at
$\lambda_{int}=+0.0048$, respectively),
(b) Goldstone mode (dot-dashed) and interaction eigenmode (dotted
curve) for the phase compared with spatial MAW profile of the phase 
gradient (full curve) in $L=2P$.
(c,d) Space time plots showing the attraction of two periods of the
same MAW
initially perturbed by the interaction eigenmode.
Parameters are $c_1=0.7, c_3=2, P=25$ near $L_3$ for (a-c) and
$c_1=3, c_3=0.85, P=21$ near $L_1$ for (d).
}\label{spec3int}
\end{figure}

An approximate prediction for the location of the apparent phase to
defect chaos transition (numerically obtained from the defect density)
can be achieved in terms of a simple linear stability analysis of the
MAWs (Figs.~\ref{specspl} and~\ref{spec3int}).  A key element in our
framework is the ``typical large value'' of $p$ as a function of
coefficients $c_1$ and $c_3$; below we will identify two linear
instabilities that act to either increase or decrease $p$, and their
balance sets a scale for typical $p$ that will predict the location of
the transition from phase to defect chaos rather well.

Due to translational and phase symmetries both MAW branches have neutral modes,
{\it i.e.,} Goldstone modes. The eigenvalue associated with the saddle-node
bifurcation is positive for MAWs of the upper branch and negative for the lower
branch. In what follows the lower branch MAWs are considered exclusively.

{\em Splitting - } The spatial structure of a MAW of large period
consists, roughly, of a homogeneous plane wave part and a local peak
part.  For the parameter regime we consider here, fully extended plane
waves are linearly unstable, and so we may expect that the MAW
spectrum will be dominated by this instability for sufficiently large
values of $P$.  Our linear stability analysis indeed shows that for
appropriate parameters ($L=P$) and small enough $P$, all eigenvalues
$\lambda_i \!<\!0$, but when we increase $P$, MAWs become linearly
unstable ($\lambda_{split}\!>\!0$, Fig.~\ref{specspl}).  The shape of
the unstable eigenmodes (Fig.~\ref{specspl}b) suggests that this
instability leads to the growth of a new peak in the homogeneous part
of the MAW, and this is indeed the behavior observed in numerical
simulations of the perturbed MAW (Fig.~\ref{specspl}c,d).  As a result
two (or more) short MAWs with smaller $P$ will appear.  We interpret
this process as the {\em splitting} of a MAW in two or more smaller
MAWs and we call the eigenmodes associated to such instability
``splitting modes''.

Clearly, this instability tends to reduce the peak-to-peak distances
$p$ and prevents MAWs to cross the SN boundary; in the phase chaotic
regime this instability {\em tends to inhibit defect generation}.

{\em Interaction - } By using a Bloch Ansatz \cite{bloch}, we extended
the stability analysis to systems with $n$ identical pulses ($L\! =\!
nP$). For $n\!>\!1$, an additional instability may appear
\cite{michal} (see Fig.~\ref{spec3int}).  Eigenvalues
$\lambda_{int}\!>\!0$ are found mainly for small $P$ (typically $P <
30$). The shape of the eigenmodes, {\it i.e.,} an alternating sequence
of positive and negative translational Goldstone modes
(Fig.~\ref{spec3int}b), suggests that the instability is due to the
{\em interaction} between adjacent MAWs. This interaction shifts
adjacent peaks into opposite directions, thereby creating occasional
larger values of $p$ (Fig.~\ref{spec3int}c,d). In phase chaos this
process leads to an increase of the spacing $p$ between some peaks,
thus {\em enhancing the generation of defects}.

{\em Competition of Instabilities - } Both the splitting and
interaction mechanisms are similar to instabilities observed in the
Kuramoto-Sivashinsky equation \cite{KS,chang}.  We believe that phase
chaos is governed by the competition of these two mechanisms that tend
to increase or decrease the inter-peak spacings $p$.  Almost
independent of the coefficients the splitting instability dominates
for MAWs with $P>30$. This can explain why large inter-peak spacings
$p>30$ become rare as reported in Figs.~\ref{distL1},\ref{distL3}.

We suggest a connection between the interchanging
dominance of these two different instabilities and the sudden change
of $\delta_D$ (near $L_1$) or the transient times before defect
occurrence (near $L_3$). We calculated the linear stability spectra
for a variety of coefficients and periods $P$ close to $P_{SN}$. From
these we obtain a curve in coefficient space (Fig.~\ref{bounds}) where
the real parts of interaction and splitting eigenvalues are {\em
equal}. For larger $c_1$ or $c_3$, interaction becomes stronger, and
we expect larger $p$'s and defect formation, while for smaller $c_1$
and $c_3$, splitting dominates, $p$'s are decreased and defect
formation becomes rare.

As shown in Fig.~\ref{bounds}, the curve where the two instabilities
are equally strong near the saddle-node bifurcation gives a rather
good estimate of where the apparent transition from phase to defect
chaos occurs. Notice that in this ``balance of instabilities''
picture, there is no tunable parameter: once we have calculated
$P_{SN}$ and the instabilities of the MAWs for a range of
coefficients, a precise prediction for the ``transition'' from phase
to defect chaos can be given.

\section{Discussion and Final Remarks}\label{conclu}

In this section we report some open questions related to defect
formation, together with some final remarks and a brief outlook.

{\em Further Refinements -} In order to accurately test our results,
we have measured for each of the 17 cuts and for several values of the
coefficients across the $L_1$- or $L_3$-lines the amplitude
distribution $D(|A|)$ and the phase gradient peak-to-peak spacing
distribution $D(p)$.  We conjectured that defects occur if and only if
$p > P_{SN}$. Indeed, we observe that in 11 out of 17 points such
conjecture is fulfilled. On the remaining 6 points the theoretical
conjecture leads to an estimation of the transition lines within a
maximal error bar of $3$\%.  The points determined following the
conjecture are indicated as empty circles in Fig.~\ref{fig1}. The
small deviations may have different reasons, that we summarize below:

\begin{figure} \vspace{0cm}
\epsfxsize=.7\hsize \mbox{\hspace*{.15 \hsize} \epsffile{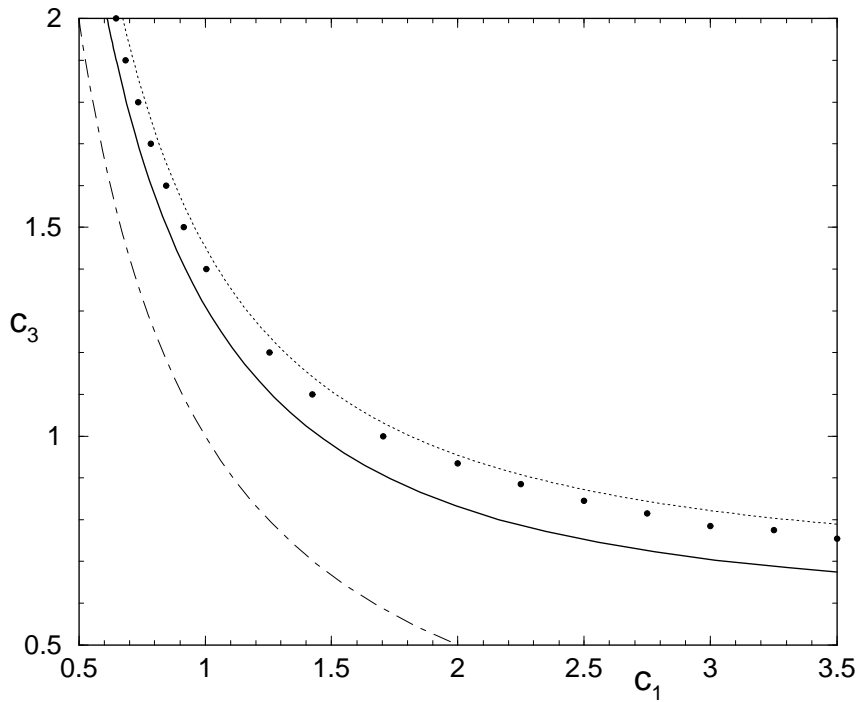}}\vspace{-1cm}
\caption[]{Space of $(c_1,c_3)$-coefficients for the CGLE with the
Benjamin-Feir-Newell curve (dot-dashed), the lower bound
$P_{SN}\!\to\!\infty$ (full) and the stability-based estimate (dotted)
for the transition from phase to defect chaos. The symbols refer to
our numerical data for the appearance of defects.}
\label{bounds}
\end{figure}

{\it (i)} If fluctuations occurring during the phase chaotic dynamics
are only moderate, such as happens near the $L_3$ transition line or
for small system sizes, more complex coherent structures can survive
for a short time.  Here we analyzed only the shortest coherent
structures characterized by a single hump.  We believe that this is
sufficient to understand the main aspects of the dynamics of large
systems. However, longer combined MAWs with more than one hump emerge
from periodic MAWs via period doubling bifurcations.  The existence of
the long combined MAWs is limited by saddle-node bifurcations
analogously to single MAWs, but these bifurcations occur at slightly
bigger values of the parameters $c_1$ and $c_3$.  Therefore the
appearance of these more complicated structures can delay defect
formation even if one inter-peak spacing within the structure is bigger
than $P_{SN}$ of the single MAW.

{\it (ii)} Near the $L_1$ line the dynamical fluctuations in the phase
chaotic regime are stronger than in the proximity of the $L_3$
line. In this case and for sufficiently high values of the parameter
$c_1$ we observed situations where not only the structure with the
longest inter-peak spacing but also the neighboring structures were
involved in the defect formation.

{\it (iii)} The assumption to consider MAWs with $\nu=0$ is only an
approximation. If the average phase gradient locally (on scales $P$)
deviates from $0$ then the saddle-node bifurcation slightly shifts
towards smaller coefficients \cite{nonzero}.

As far as the (numerically) improved $L_3$ and $L_1$ lines are
concerned, we observe that both these lines lie to the left of the
ones determined in earlier numerical studies \cite{chate}. This is due
to the fact that our simulations are of longer duration then those
performed previously.  This confirms the expectation that such
transition lines will shift towards the Benjamin-Feir-Newell curve for
increasing systemsize and integration times \cite{egolf}. Moreover,
some authors claim that indeed in the thermodynamic limit $L_1$ and
$L_3$ will coincide with the Benjamin-Feir-Newell curve and the phase
chaos regime will disappear~\cite{chate2d}. On the basis of our
simulations we cannot exclude such a possibility for higher space
dimensions, but based on the results presented in this paper we
conjecture that the saddle-node line for $P \!\rightarrow\!\infty$
provides a lower boundary for the transition from phase to defect
chaos in the one-dimensional CGLE.

{\em Final Remarks -} We have presented a systematic study of
modulated amplitude waves (MAWs) in the complex Ginzburg-Landau
equation (CGLE).  These periodic coherent structures originate from
supercritical bifurcations from the homogeneous oscillation of the
CGLE due to the Benjamin-Feir instability. The range of existence of
MAWs is bounded by saddle-node bifurcations occurring for values of
$c_1$ and $c_3$ that depend on the period $P$ of the MAWs.
Approaching the transition from phase to defect chaos, near-MAWs with
large $P$ occur in phase chaos, and defects are generated when the
period of these near-MAWs becomes larger than the spatial period
$P_{SN}$ of the critical nucleus.  This scenario is valid for both the
$L_1$ and $L_3$ transition. The divergence of $P_{SN}$ for
coefficients in the phase-chaos regime led us to conjecture that there
is a lower bound for the transition from phase to defect chaos.
Considerations of the linear stability properties of MAWs in light of
their tendency to increase or decrease the typical period $p$ in phase
chaos, has led us to a fit-free estimate of the apparent transition
from phase to defect chaos that fits the numerical data well.

Altogether, our study leaves little space for doubt that the
transition from phase chaos to defect chaos in the CGLE is governed by
coherent structures and their bifurcations.  From a general viewpoint,
our analysis shows that there is no collective behavior that drives
the transition. Instead, strictly local fluctuations drive local
structures beyond their saddle-node bifurcation and create defects.

{\em Outlook -} We want to stress here that the extension of the
analysis to MAWs with nonzero average phase gradients~\cite{nonzero},
will be of considerable interest for experimentalists, because in some
recent experiments concerning Rayleigh-B\'enard or Marangoni
convection in quasi-one-dimensional geometries, supercritical Eckhaus
instabilities of plane wave trains and the corresponding emergence of
stable saturated MAWs have been observed
\cite{SupEck,velarde,janiaud}. These states are analogous to what
happens for the 1d CGLE when phase chaotic solutions with $\nu \ne 0$
are considered \cite{miguel,torcini}.

The relevance of MAWs for two-dimensional structures is suggested by
recent experimental evidence of MAWs observed in connection with
superspiral and spiral breakup occurring in a Belousov-Zhabotinsky
reaction \cite{chem}.  Moreover, in the phase chaotic regime of the 2d
CGLE the correspondence between long inter-peak spacings (here diameter
of cells) and the strength of the local modulation has already been
noticed numerically \cite{chate2d}. Additional mechanisms present in
2d remain to be explored.  Thereby it might turn out that phase chaos
exists in the thermodynamic limit in 1d only but not in 2d as
previously conjectured \cite{chate2d}.

%\acknowledgements 

It is a pleasure to acknowledge discussions with H. Chat\'e, M. Howard
and L. Kramer. AT and MB are grateful to ISI Torino for providing a
pleasant working environment during the Workshop on ``Complexity and
Chaos'' in October 1999.  AT would also thank Caterina, Daniel,
Katharina and Sara for providing him with a faithful representation of
a chaotic evolution.  MGZ is supported from a post-doctoral grant of
the MEC (Spain) and FOMEC-UBA (Argentina). MvH acknowledges financial
support from the EU under contract ERBFMBICT 972554.


\begin{thebibliography}{9}
  
\bibitem{maw} L. Brusch, M. G. Zimmermann, M. van Hecke, M. B\"ar and
A.  Torcini, Phys. Rev. Lett. {\bf85}, 86 (2000).

\bibitem{kura} Y. Kuramoto, {\em Chemical Oscillations, Waves
and Turbulence} (Springer, 1984, Berlin).

\bibitem{CH} M. C. Cross and P. C. Hohenberg, Rev. Mod. Phys. {\bf
65}, 851 (1993).

\bibitem{KS} P. Manneville, {\em Dissipative structures and Weak Turbulence}
(Academic Press, 1990, San Diego);
T. Bohr, M. H. Jensen, G. Paladin and A. Vulpiani, {\em Dynamical
systems approach to turbulence} (Cambridge Univ. Press) (1998).

\bibitem{review} A recent and detailed review on the CGLE is
the following : I.S. Aranson and L. Kramer, {\em The World
of Complex Ginzburg-Landau Equation}, to appear in Rev. Mod. Phys.

\bibitem{wire} J. M. Vince and M. Dubois, Physica D, {\bf 102}, 93
(1997).

\bibitem{film} M. Rabaud, S. Michalland and Y. Couder,
Phys. Rev. Lett. {\bf 64}, 184 (1990); D. P. Vallette, G. Jacobs and
J. P. Gollub, Phys. Rev. E {\bf55}, 4274 (1997).

\bibitem{eutectic} S. Akamatsu and G. Faivre, 
Phys. Rev. E {\bf 58}, 3302 (1998).

\bibitem{binary} M. L\"ucke, W. Barten and M. Kamps, 
Rhysica D, {\bf 61}, 183 (1992).

\bibitem{sidewall} Y. Liu and R. E. Ecke, Phys. Rev. Lett. {\bf 78},
4391 (1997).

\bibitem{chem} Q. Ouyang and J. M. Flesselles, Nature {\bf 379}, 143
(1996); Q. Ouyang, H. L. Swinney and G. Li, Phys. Rev. Lett. {\bf 84},
1047 (2000); L. Q. Zhou and Q. Ouyang, Phys. Rev. Lett. {\bf 85}, 1650
(2000).

\bibitem{dean} P. Bot and I. Mutabazi, Eur. Phys. J. B {\bf 13}, 141 (2000).

\bibitem{SupEck} N. Mukolobwiez, A. Chiffaudel and F. Daviaud, 
Phys. Rev. Lett. {\bf 80}, 4661 (1998);
J. Burguete, H. Chat\'e, F. Daviaud and N. Mukolobwiez, 
Phys. Rev. Lett. {\bf 82}, 3252 (1999).

\bibitem{velarde} 
A. Wierschem, H. Linde and M. G. Velarde, Phys. Rev. E {\bf 62}, 6522
(2000).

\bibitem{janiaud} B. Janiaud, A. Pumir, D. Bensimon, V. Croquette, H.
  Richter and L. Kramer, Physica D, {\bf 55}, 269 (1992).

\bibitem{pumir}  A. Pumir, B. I. Shraiman, W. van Saarloos,
P. C. Hohenberg,  H. Chat\'e and M. Holen, p. 173 in
C. D. Andereck and F. Hayot (Eds.),``Ordered and Turbulent patterns in
Taylor-Couette Flow'' (Plenum Press, New York, 1992)

\bibitem{chao1}  B. I. Shraiman,  A. Pumir,  W. van Saarloos,
P. C. Hohenberg,  H. Chat\'e and M. Holen, Physica D {\bf 57}, 241
(1992). 

\bibitem{bazhenov} M. V. Bazhenov, M. I. Rabinovich and A. L. Fabrikant, 
Phys. Lett. A {\bf 163}, 87 (1994). 

\bibitem{chate} H. Chat\'e, Nonlinearity {\bf 7}, 185 (1994); p. 33 in
 P. E. Cladis and Palffy-Muhoray (Eds.),``Spatio-Temporal Pattern Formation in
 Nonequilibrium Complex Systems'' (Addison Wesley, Reading, 1995).

\bibitem{saka} H. Sakaguchi, Prog. Theor. Phys. {\bf 84}, 792 (1990).

\bibitem{egolf} D. A. Egolf and H. S. Greenside, Phys. Rev. Lett.
  {\bf74} 1751 (1995)

\bibitem{miguel} R. Montagne, E. Hern\'andez-Garc\'{\i}a and M. San Miguel,
Phys. Rev. Lett. {\bf 77}, 1047 (1996); R. Montagne,
E. Hern\'andez-Garc\'{\i}a, A. Amengual and M. San Miguel, Phys. Rev. E {\bf
55}, 151 (1997).

\bibitem{torcini} A. Torcini, Phys. Rev. Lett. {\bf 77}, 1047 (1996);
 A. Torcini, H. Frauenkron and P. Grassberger, Phys.
  Rev E {\bf 55}, 5073 (1997).

\bibitem{saar} W. van Saarloos and P. C. Hohenberg, Physica D {\bf
56}, 303 (1992); {\bf 69}, 209 (1993) [Errata].

\bibitem{mvh} M. van Hecke, Phys. Rev. Lett. {\bf 80}, 1896 (1998).

\bibitem{hegger} G. Giacomelli, R. Hegger, A. Politi and M. Vassalli, 
Phys. Rev. Lett. {\bf 85}, 3616 (2000).

\bibitem{chate2d} P. Manneville and H. Chat\'e, Physica D {\bf
96}, 30 (1996).

\bibitem{note1} By substituting $\kappa\!:=\!a_z/a$ one reproduces the form of
the ODEs used in \cite{saar} which is more appropriate for studies of fronts.

\bibitem{nb} K. Nozaki and N. Bekki, J. Phys. Soc. Jpn.
{\bf 53} (1984) 1581.

\bibitem{notemulti} For completeness we point out that the ODEs 
(\ref{ode}) also contain complicated multi-loop orbits that correspond 
to more complex coherent structures which have a small basin of 
attraction and little relevance for the dynamics of the CGLE.

\bibitem{mm} M. van Hecke and M. Howard, Phys. Rev. Lett. {\bf 86},
2018 (2001).

\bibitem{hager} G. Hager, {\it Quasiperiodische L\"osungen der eindimensionalen
komplexen Ginzburg-Landau Gleichung}, Diploma Thesis, University of Bayreuth,
Germany (1996).

\bibitem{note3} The maximal ``conserved'' (during time evolution) 
average phase gradient vanishes approaching the transition to
defect chaos~\cite{miguel,torcini}. This result is rigorous only on scales 
of the system size but for smaller portions $\nu$ can fluctuate around 0. 
Typically we observe quasi-coherent structures (near-MAWs) in the phase chaotic 
regime with associate $\nu$-values in the interval $[-0.01,+0.01]$. 
MAWs with such small $\nu$ do not deviate much from the $\nu=0$ MAWs
\cite{nonzero}, therefore the comparison of the observed structures with the
$\nu=0$ MAWs is satisfactory.

\bibitem{nonzero} L. Brusch, A. Torcini and M. B{\"a}r, in
preparation.

\bibitem{Auto94} E. J. Doedel, X. J. Wang and T. F. Fairgrieve, {\em AUTO94:
Software for continuation and bifurcation problems in ordinary differential 
equations}, 
Applied Mathematics Report, California Institute of Technology (1994).

\bibitem{note2} The linear stability analysis of the
uniformly oscillating solutions $A_0(x,t) = {\rm e}^{ic_3t}$,
can be performed by considering the following perturbed solution 
$A(x,t) = (1+a(x,t)){\rm e}^{ic_3t}$ \protect\cite{KS}. Where
$a(x,t) = \sum_k a_k {\rm e}^{ikx + \lambda_k t}$ with
$\lambda_k = \sigma_k + i \Omega_k$. It is straightforward to show that
the real part of the growth-rate $\sigma_k$ is, up to fourth order in $k$,
given by
\begin{eqnarray}
\sigma_k \sim  -(1-c_1 c_3) k^2 - \frac{1}{2} (1+c_3^2) c_1^2 k^4 ~.
\nonumber
\end{eqnarray}
From this expression it is clear that $\sigma_k > 0$ only if $c_1 c_3
> 1$ (this is nothing else than the Newell criterion).  Moreover there
exists a critical value $k_0^2 = [2(c_1c_3-1)]/[c_1^2(1+c_3^2)]$ above
which $\sigma_k$ is always negative. For finite size systems the
smallest allowed wavevector is $k_{min}=2\pi/L$, therefore the uniform
oscillation becomes unstable for $k_{min} \le k_0$ and this condition
allows to derive the corresponding critical values of the parameters
$c_1$ and $c_3$.

\bibitem{DwightDP}
M. Kness, L. Tuckerman and D. Barkley, Phys. Rev. A {\bf 46}, 
5054 (1992). 

\bibitem{ioos} See chapter 3 of G. Iooss and D. D. Joseph,
{\it Elementary Stability and Bifurcation Theory}
(Springer, Berlin, 1980).

\bibitem{note_nm} An exception on this rule occurs for large $c_1$,
where the dependence of $c_3$ on $P$ at the saddle-node becomes
non-monotonic.

\bibitem{chang} H.-C. Chang, E. A. Demekhin and E. Kalaidin, SIAM J. Appl.
Math. {\bf 58}, 1246 (1998); H.-C. Chang, E. A. Demekhin and D. I. Kopelevich, 
Physica D {\bf 63}, 299 (1993).

\bibitem{abel} M. Abel, H. Chat{\'e} and H. Voss, to be published. 

\bibitem{note4} Another relevant line that appears in the parameter
plane is the so-called $L_2$-line which is the transition from defect
to phase chaos in the hysteretic regime.

\bibitem{bloch} 
N. W. Ashcroft and N. D. Mermin, {\it Solid State Physics} (Holt, Rinehart and
Winston, New York, 1976);
P. Collet and J.-P. Eckmann, {\it Instabilities and Fronts in Extended Systems}
(Princeton University Press, 1990).

\bibitem{michal} M. Or-Guil, I. G. Kevrekidis and M. B\"ar, Physica D {\bf
135}, 154 (2000).

\end{thebibliography}
\end{document}